\documentclass[ twocolumn,
aps,prd,   
               preprintnumbers,numbers,sort&compress,nofootinbib,
                            showpacs,
               colorlinks,
               linkcolor=blue,   
               citecolor=blue]{revtex4-1}
    \newcommand{\exclude}[1]{}

\usepackage{graphicx,amsmath,amssymb,bm}
\usepackage{psfrag}
\usepackage{feynmp}
\usepackage{hyperref}
\usepackage{slashed}
\usepackage{subcaption}

\def\<{\langle}
\def\>{\rangle}

\def\+{\dagger}

\def\U1A{U(1)$_{\rm A}$}

 \def\<{\langle}
\def\>{\rangle}

\def\+{\dagger}

\def\ra{\rangle}
\def\la{\langle}
\def\U1A{U(1)$_{\rm A}$}

\newcommand{\be}{\begin{eqnarray}}
\newcommand{\ee}{\end{eqnarray}}
\newcommand{\beq}{\begin{equation}}
\newcommand{\eeq}{\end{equation}}

\graphicspath{{figures/}}
\begin{document}

\title{Cosmological   Axion  and   Quark Nugget Dark Matter Model}
%
%

\author{ Shuailiang Ge}
\author{ Xunyu Liang}
\author{Ariel Zhitnitsky}
\affiliation{Department of Physics and Astronomy, University of
  British Columbia, Vancouver,  Canada}
  
\begin{abstract}
    
We study  a dark matter (DM) model offering a very natural  explanation of  two  (naively unrelated) problems   in cosmology: the observed relation $\Omega_{\rm DM}\sim\Omega_{\rm visible}$ and the observed asymmetry between matter and antimatter in the Universe, known as the ``baryogenesis" problem. In this framework, both types of matter (dark and visible) have the same QCD origin, form at the same QCD epoch, and both proportional to one and the same dimensional parameter of the system, $\Lambda_{\rm QCD}$, which  explains how these two, naively distinct,  problems could be  intimately related, and could be solved simultaneously within the same framework. 
 More specifically, the DM in this model is composed by two different ingredients: the (well- studied) DM axions and  (less-studied) the quark nuggets made of matter or antimatter.     The focus of the present work is the quantitative analysis  of the relation between these two distinct  components contributing to the dark sector of the theory determined by $\Omega_{\rm DM}\equiv [\Omega_{\rm DM}(\rm nuggets)+ \Omega_{\rm DM}(\rm axion)]$.   
We argue that  the  nugget's DM component  always traces the visible matter density, i.e. $\Omega_{\rm DM}(\rm nuggets)\sim\Omega_{\rm visible}$ and this feature   is  not sensitive to  the parameters of the system such as the axion mass $m_a$ or  the misalignment angle $\theta_0$.
 It should be contrasted with conventional axion production mechanism due to the misalignment   when $\Omega_{\rm DM}(\rm axion)$   is highly sensitive to  the axion mass $m_a$ and  the  initial  misalignment angle $\theta_0$. We also discuss the  constraints on this model related to the inflationary scale $H_I$, non-observation of the isocurvature perturbations  and the tensor modes. We also comment on some constraints  related to  varies  axion search experiments.  
  
\end{abstract}
\vspace{0.1in}

\maketitle
\section{Introduction and Motivation}
\label{intro}

The idea that the dark matter may take the form of composite objects of 
standard model quarks in a novel phase goes back to quark nuggets  \cite{Witten:1984rs}, strangelets \cite{Farhi:1984qu}, nuclearities \cite{DeRujula:1984axn},  see also review \cite{Madsen:1998uh} with large number of references on the original results. 
In the models \cite{Witten:1984rs,Farhi:1984qu,DeRujula:1984axn,Madsen:1998uh}  the presence of strange quark stabilizes the quark matter at sufficiently 
high densities allowing strangelets being formed in the early universe to remain stable 
over cosmological timescales.  There were a number of problems with the original idea\footnote{\label{first-order}In particular, the first order phase transition is a required feature of the system for the strangelet  to be formed during the QCD phase transition.  However it is known by now that the QCD transition is a crossover rather than the first order phase transition as the recent lattice results \cite{Aoki:2006we} unambiguously show. Furthermore, the strangelets 
will likely evaporate on the Hubble time-scale even if they had been formed \cite{Alcock:1985}.} and we refer to the review paper \cite{Madsen:1998uh} for the details. 

The quark nugget model advocated in \cite{Zhitnitsky:2002qa} is conceptually similar, with the 
nuggets being composed of a high density colour superconducting (CS) phase.
An additional stabilization factor in the quark nugget model is provided by the axion domain walls
  which are copiously produced during the QCD transition\footnote{In this case the first order phase transition is not required for the nuggets to be formed as the axion domain wall plays the role of the squeezer. Furthermore, the argument  related to the fast evaporation of the strangelets as mentioned in  footnote \ref{first-order}   is not applicable for the  quark nugget model \cite{Zhitnitsky:2002qa} because the  vacuum ground state energies inside (CS phase) and outside (hadronic phase)  the nuggets are drastically different. Therefore these two systems can coexist only in the presence of the additional external pressure provided by the axion domain wall, in contrast with strangelet models \cite{Witten:1984rs,Madsen:1998uh} which  must  be stable at zero external pressure.}.
 The only new additional crucial element in the proposal \cite{Zhitnitsky:2002qa}   is that the nuggets could be 
made of matter as well as {\it antimatter} in this framework, see original papers \cite{Zhitnitsky:2002qa, Oaknin:2003uv}.

This novel  crucial element of the model \cite{Zhitnitsky:2002qa, Oaknin:2003uv} completely changes entire framework 
because the dark matter density  $\Omega_{\rm dark}$ and the baryonic matter density $ \Omega_{\rm visible}$ now become intimately related
to each other and proportional to each other
\be
\label{Omega}
 \Omega_{\rm dark}\sim \Omega_{\rm visible}
\ee
as will be explained  in next section \ref{sec:AQN}. In other words, the nature of dark matter and the problem of the asymmetry between matter and antimatter in the Universe, normally formulated as the so-called baryogenesis problem,   become  two sides of the same coin in this framework. To reiterate this claim: the relation   (\ref{Omega})  is  very generic outcome of the framework, and it is  not sensitive to any specific details of the model.  

The proposal \cite{Zhitnitsky:2002qa, Oaknin:2003uv} represents an 
alternative to baryogenesis  scenario  when  the 
``baryogenesis'' is actually a charge separation process 
in which the global baryon number of the Universe remains 
zero. In this model the unobserved antibaryons  come to comprise 
the dark matter in the form of dense antinuggets in colour superconducting (CS) phase.  The dense nuggets in CS phase also present in the system
such that the total baryon charge remains zero at all times during the evolution of the Universe. The detail mechanism of the formation of the  nuggets 
and antinuggets has been recently developed in refs.   \cite{Liang:2016tqc,Ge:2017ttc}, and the present  work can be considered as a continuation 
of these studies. 
We review the basics elements of this proposal, its predictions and the observational consequences including presently available constraints,     in next Section \ref{sec:AQN}.  We highlight in few paragraphs below
the main questions which will be    addressed in the present work.

Our first paper \cite{Liang:2016tqc} devoted to the formation of the nuggets was focused on the dynamics of a single nugget, its formation, and its time evolution. If the Universe were symmetric during the formation time, the equal number of nuggets and antinuggets would   form. However, the coherent 
   (on the enormous scale of the Universe) $\cal{CP}-$ odd axion field was present in the system. Precisely this asymmetry generates the preferential evolution of the system such that number of nuggets and antinuggets would  not be  identically the same, which was the main result of  our  paper \cite{Ge:2017ttc}. As the total baryon charge is conserved in the system, the remaining baryon charge (not hidden in form of the   nuggets and antinuggets)   represents the visible baryonic matter  contributing to $ \Omega_{\rm visible}$. This represents a qualitative explanation of   generic consequence of this entire framework  as represented by   (\ref{Omega}).

However, in the   previous papers \cite{Liang:2016tqc,Ge:2017ttc} our main objective was to understand a novel  fundamental phenomenon  on a qualitative level by  simplifying  the system itself and neglecting a large number of specific features of the system which may produce quantitative effects of order one, but cannot change the qualitative picture advocated in \cite{Liang:2016tqc,Ge:2017ttc}. 

Indeed, in our previous studies we ignored a number of  effects. In particular:  a)   We assumed that the energy per baryon charge is the same for hadronic   and CS phases; b) We ignored  the contribution of the axion domain wall surrounding the quark nugget in CS phase using a simple, order of magnitude estimate for the mass of the nugget, $M_{\rm nugget}\sim m_p B$; c)We neglected the contribution of  the conventional  propagating axions
produced by misalignment mechanism (or by decay of the topological defects)    by  assuming that  $\Omega_{\rm dark}$  is saturated by the nuggets.
All these simplifications allowed us to  relate  $ \Omega_{\rm visible}$ and   $\Omega_{\rm dark}$ in terms of a single parameter $c$ defined as a ratio of the  hidden baryon charge in  nuggets and antinuggets, see precise definition in eq.  (\ref{eq:c_previous}).

The main  goal of the present work is to take into account all  these complications and numerical factors and  incorporate  them into a new equation which would be    generalization of eq.  (\ref{eq:c_previous}).  We think that the timing for this task  is quite appropriate  due to a number of reasons. First, there is number of groups searching for the axion, see relevant references in next section. If the axion is found with definite mass $m_a$ it may not necessary saturate the entire 
dark matter density $\Omega_{\rm DM}$.   A finite portion of $\Omega_{\rm DM}$ could be related to the nuggets contribution which, according to  
refs. \cite{Liang:2016tqc,Ge:2017ttc}, always accompanies the conventional axion production.  We refer to the  original papers \cite{axion,KSVZ,DFSZ}  and  recent reviews   
  \cite{vanBibber:2006rb, Asztalos:2006kz,Sikivie:2008,Raffelt:2006cw,Sikivie:2009fv,Rosenberg:2015kxa,Marsh:2015xka,Graham:2015ouw,Ringwald2016} on the axion dynamics and recent axion searches. 
  
Our goal is to relate these two complementary mechanisms of the axion dynamics:  some portion of the axion field will  radiate the propagating axions, while another portion of the coherent axion field will lead to the nugget's formation. These two mechanisms accompany each other and we want to establish a quantitative relation between the two.

Secondly, it has been recently claimed that the non-observation of the isocurvature perturbations  and the tensor modes impose strong constraints on relation between the axion parameter $f_a$ and inflation scale $H_I$. Our comment here is that such kind of constraints are normally imposed by   assuming  that the relic  axions saturates the DM density. 
However,  if the nuggets become the dominant contributor to $\Omega_{\rm DM}$, the corresponding constraint   drastically weakens  because the axion's contribution   to $\Omega_{\rm DM}$ becomes subdominant, and  as a consequence,   the  isocurvature perturbations will be suppressed for the same axion  mass $m_a$ and  coupling  $f_a$. These few  comments suggest  that there is a number of relevant parameters in the theory which crucially depend on relative contribution of the relic axions versus  nugget's portion  to $\Omega_{\rm DM}$. The goal of this work is to elaborate on those relations.

The paper is organized as follows. In section \ref{sec:AQN} we overview the basic idea and phenomenological consequences on the proposal   \cite{Zhitnitsky:2002qa, Oaknin:2003uv} with emphasis how the ``baryogenesis" is replaced by the charge separation effect  \cite{Liang:2016tqc,Ge:2017ttc}. 
In section \ref{sec:2.density}, we will introduce our notations and definitions of the parameters to be computed.  In section \ref{sec:3.epsilon model}, we will focus on  the internal structure of the nuggets. In particular, we compute  the energy per baryon charge of a single stabilized (anti)nugget  at zero temperature. Then in section \ref{sec:4.c}, using the results of previous two sections, we produce a number of numerical  plots and show that there is a large region of parametrical space in this model which is   consistent with all presently available constraints. The corresponding constraints have been obtained  from a variety of independent astrophysical, cosmological, satellite and ground based observations.  Furthermore,  we will show that this model is also consistent with known constraints from the axion search experiments. In section \ref{sec:5.conclusion}, we briefly summarize the main results of this paper and discuss the potential future directions. In particular, we speculate that unexpected correlations between the frequency of appearance of the solar flares and the intensity of the dark matter fluxes  observed in the Sun, might be related to the antiquark nuggets studied in this work.

 \section{Axion Quark Nugget (AQN)  model}
\label{sec:AQN}

The AQN Model in the title of this section stands for the axion quark nugget   model to emphasize on essential role of the axion field and avoid confusion with earlier models such as quark nuggets, strangelets, nuclearities mentioned in the Introduction. 
\exclude{The AQN can    also stand for  as  anti-quark nugget because the anti-nuggets can only appear in the   system as a result of the coherent  $\cal{CP}$ odd axion field which  separates   the baryon charges as
discussed in \cite{Liang:2016tqc,Ge:2017ttc}   and will be reviewed  below. The antinuggets  cannot appear in   other models mentioned in Introduction.
}

The basic ideas of  the  original proposal \cite{Zhitnitsky:2002qa, Oaknin:2003uv} can be explained   as follows: 
It is commonly  assumed that the Universe 
began in a symmetric state with zero global baryonic charge 
and later (through some baryon number violating process, so-called ``baryogenesis") 
evolved into a state with a net positive baryon number. As an 
alternative to this scenario we advocate a model in which the 
baryogenesis is actually a charge separation process 
in which the global baryon number of the universe remains 
zero. In this model the unobserved antibaryons come to comprise 
the dark matter in the form of dense nuggets of quarks and antiquarks in CS phase.  
  The formation of the  nuggets made of 
matter and antimatter occurs through the dynamics of shrinking axion domain walls, see
original papers \cite{Liang:2016tqc,Ge:2017ttc} with the technical details. 

\begin{figure}
\centering
\captionsetup{justification=raggedright}
\includegraphics[width=0.8\linewidth]{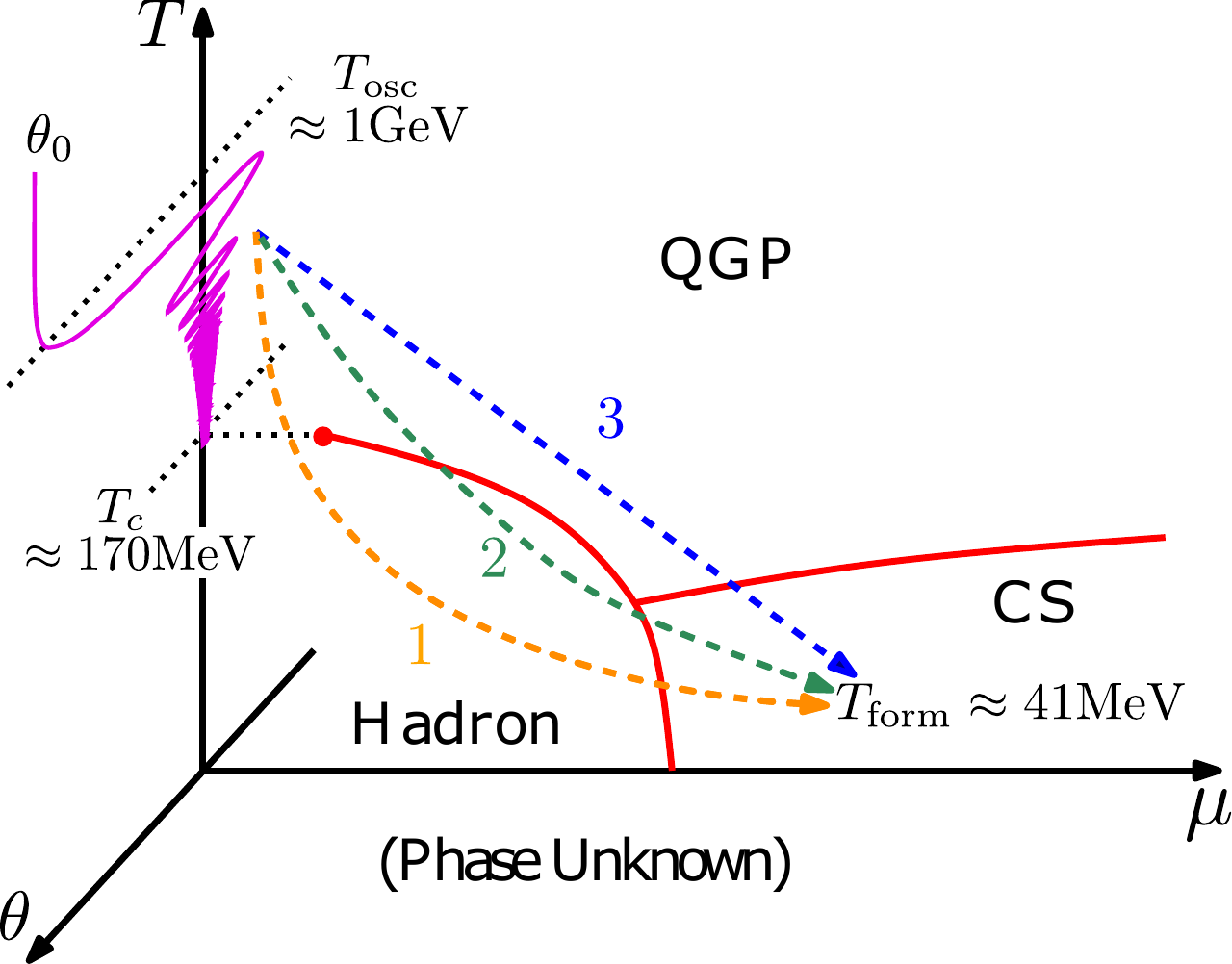}
\caption{This diagram illustrates the interrelation between the axion production due to the misalignment mechanism and the nugget's formation which starts before the axion field $\theta$ relaxes to zero. Possible cooling paths are denoted as path 1, 2 or  3. The phase diagram is in fact much more complicated as the dependence on the third essential parameter, the $\theta$ is not shown as it is largely unknown. It is assumed that the initial axion field starts at $\theta_0$, while   the nuggets complete their formation in  the CS region  at  $T_{\rm form}\approx 41$ MeV, $\mu >\mu_c$ and $\theta\approx 0$.   }
\label{phase_diagram}
\end{figure}

If the fundamental $\theta$ parameter of QCD were identically zero  
during the formation time, equal numbers of nuggets 
made of matter and antimatter would be formed.  However, the fundamental $\cal CP$ violating processes associated 
with the $\theta$ term in QCD   result in the preferential formation of 
antinuggets over nuggets.   This source of strong $\cal CP$ violation is no longer available at the present epoch as a result of the axion dynamics  when $\theta$ eventually relaxes to zero, see Fig. \ref{phase_diagram}. 
     As a result of these $\cal CP$ violating processes the number of nuggets and antinuggets 
      being formed would be different. This difference is always of order of one effect   irrespectively to the parameters of the theory, the axion mass $m_a$ or the initial misalignment angle $\theta_0$, as argued in  \cite{Liang:2016tqc,Ge:2017ttc}.   As a result of this disparity between nuggets and antinuggets   a similar disparity would also emerge between visible quarks and antiquarks.  Precisely this disparity between visible baryons  and antibaryons eventually lead (as a result of the annihilation processes) to the system when exclusively one  species of visible baryons  remain in the system,  in agreement with observations. 
      
      We illustrate this interrelation between the axion dynamics and the nugget's formation on Fig.  \ref{phase_diagram} when both processes occur during the same QCD epoch. The key element of the proposal is the presence of the coherent axion $\theta(t)$ field  which continues to oscillate when the nugget's formation starts.   
  The time evolution of the nuggets        continues until low temperature $T_{\rm form}$ when the axion field assumes its final destination $\theta=0$ which is observed today. However, the originally produced  asymmetry between nuggets and antinuggets    cannot be washed out  as explained in \cite{Ge:2017ttc}.

      This disparity between nuggets and antinuggets  unambiguously implies that  the total number of  antibaryons  will be less than the number of baryons in   early universe plasma.
 This  is precisely  the reason why the resulting visible and dark matter 
densities must be the same order of magnitude (\ref{Omega})  in this framework  
as they are both proportional to the same fundamental $\Lambda_{\rm QCD} $ scale,  
and they both are originated at the same  QCD epoch.
  If these processes 
are not fundamentally related, the two components 
$\Omega_{\rm dark}$ and $\Omega_{\rm visible}$  could easily 
exist at vastly different scales.

 Another fundamental ratio 
is the baryon to entropy ratio at present time
\be
\label{eta}
\eta\equiv\frac{n_B-n_{\bar{B}}}{n_{\gamma}}\simeq \frac{n_B}{n_{\gamma}}\sim 10^{-10}.
\ee
In our proposal (in contrast with conventional baryogenesis frameworks) this ratio 
is determined by the formation temperature $T_{\rm form}\simeq 41 $~MeV  at which the nuggets and 
antinuggets complete their formation.  We note that $T_{\rm form}\approx \Lambda_{\rm QCD}$. This temperature    is determined by the observed  ratio (\ref{eta}). The $T_{\rm form}$  assumes a typical QCD value, as it should as there are no any small parameters in QCD, see Fig. \ref{phase_diagram}.

Unlike conventional dark matter candidates, such as WIMPs 
(Weakly Interacting Massive Particles) the dark-matter/antimatter
nuggets are strongly interacting and macroscopically large, nuclear density 
objects with a typical size $(10^{-5}-10^{-4})$ cm, and  the baryon charge which 
ranges from $B\sim 10^{23}$ to $B\sim 10^{28}$. However, they do not contradict any of the many known observational
constraints on dark matter or
antimatter  for three main reasons~\cite{Zhitnitsky:2006vt}:
\begin{itemize}
	\item They carry very large baryon charge $|B|>10^{23}$ which is determined by the size of the nugget $\sim m_a^{-1}$. As a result, 
	  the number density of the nuggets is very small $\sim B^{-1}$. Their interaction with visible matter is thus highly suppressed. In particular, the quark nuggets essentially decouple from cosmic microwave background (CMB) photons, and therefore, they do not destroy conventional picture for the structure formation;
	\item The nuggets has a huge mass $M_{\rm nugget}\sim m_pB$, therefore the effective interaction is very small $\sigma/M_{\rm nugget}\sim10^{-10}{\rm cm}^2/{\rm g}$, which is evidently well below the upper limit of the conventional DM constraint $\sigma/M_{\rm DM}<1{\rm cm}^2/{\rm g}$;
	\item The quark nuggets have very large binding energy due to the large gap $\Delta\sim100$ MeV in the CS phase. Therefore, the strongly bound baryon charge is unavailable to participate in the big bang nucleosynthesis (BBN) at $T\approx1{\rm MeV}$, long after the nuggets had been formed.
\end{itemize}
 We emphasize that the weakness of the visible-dark matter interaction 
in this model is due to a  small geometrical parameter $\sigma/M \sim B^{-1/3}$ 
  which replaces 
the conventional requirement of sufficiently weak interactions for WIMPs. 

In this short overview of the model we also  want to make few comments regarding the 
observational consequences of the proposal. 
  It is known that  the galactic spectrum 
contains several excesses of diffuse emission the origin of which is uncertain, the best 
known example being the strong galactic 511~keV line.  Our comment here is that this model    offers a 
potential explanation for several of 
these diffuse components (including 511 keV line and accompanied   continuum of $\gamma$ rays in 100 keV and few  MeV ranges, 
as well as x-rays,  and radio frequency bands). For further details see the original works 
\cite{Oaknin:2004mn, Zhitnitsky:2006tu,Forbes:2006ba, Lawson:2007kp,
Forbes:2008uf,Forbes:2009wg,Lawson:2012zu}   with specific computations in different frequency bands in galactic radiation, and a short overview
\cite{Lawson:2013bya}.  
\exclude{
To reiterate: this model  is consistent with all known astrophysical, cosmological, satellite and ground based constraints as highlighted above. 
Furthermore, in a number of cases the predictions of the models are very close to the presently available limits, and very modest improving of those constraints may lead to a discovery of the nuggets. Even more than that: there is a number of frequency bands where some excess of emission was observed, and this model may explain some portion, or even entire excess of the observed radiation in these frequency bands. 
}

  As we already mentioned the  typical size of  a nugget $\sim m_a^{-1}$ is determined by the axion mass, yet to be discovered. Therefore, a 
 typical  baryon charge of the nuggets and their size distribution are also unknown characteristics  of the model.  However, it is important to emphasize that the   allowed window for the axion mass $10^{-6} {\rm eV}\leq m_a \leq 10^{-2} {\rm eV}$
 corresponds to the range of the nugget's baryon charge $B$ which   largely overlaps  with all presently available and independent constraints on such kind of dark matter masses and baryon charges 
 \beq
 \label{B-range}
 10^{23}\leq |B|\leq 10^{28}, 
 \eeq
 see e.g. \cite{Jacobs:2014yca,Lawson:2013bya} for review\footnote{\label{B-constraint}The smallest nuggets with $B\sim (10^{23}-10^{24})$ naively  contradict to the constraints cited in \cite{Jacobs:2014yca}. However, the corresponding constraints are actually derived with the assumption that   nuggets with a definite mass (smaller than 55g) saturate the dark matter density. In contrast, we assume  that the peak  of the nugget's distribution corresponds to a larger value of mass, $\la B\ra \geq 10^{25}$, while the small nuggets represent a tiny portion of the total dark matter density. The same comment also applies to   the larger masses excluded by Apollo data as reviewed in  \cite{Lawson:2013bya}.
  Large nuggets with $B\sim 10^{28}$ may  exist, but represent  a small portion of the total dark matter density, and therefore, do not contradict the Apollo's constraints, see also some comments  on the baryon charge distribution  (\ref{distribution})  in concluding section \ref{sec:5.conclusion}.}. To reiterate the same claim:  the allowed window (\ref{B-range}) for the baryon charge $B$  is   perfectly   consistent with all presently available constraints. 
  
  Furthermore, if one takes the recent conjecture  \cite{Zhitnitsky:2017rop,Zhitnitsky:2018mav} 
that  the extreme ultra violet (EUV) and soft x-ray emission from the solar Corona are due to the antiquark nuggets hitting  the Sun from outer space,
then one can make one step further and identify  the observed ``nano-flares"   with annihilation events of the antinuggets in the solar corona. Taken this identification literally one can translate the observed 
 energy distribution of the ``nano-flares"   in solar corona   in terms of the nugget's baryon charge distribution. This identification leads  to an independent  estimate on possible values for $B$ which is  perfectly consistent with the allowed  range  (\ref{B-range}) for the baryon charge of the nuggets  \cite{Zhitnitsky:2017rop,Zhitnitsky:2018mav}. Therefore, in our analysis     which follows we treat the window (\ref{B-range}) as the solid constraint on the allowed magnitude of  the nugget's baryon charge. 

     \section{The AQN Model. Notations and Definitions}
\label{sec:2.density}

In this section, we will first briefly introduce our notations for the AQN  model which  is constituted by two parts: one portion  is  represented by the nuggets and antinuggets, while  the other component  is represented by free propagating axions which were produced  by the  axion field (through e.g.  the misalignment mechanism). In order  to quantitatively describe this model, we have to  derive a general expression  which relates these two complimentary contributions to  the $\Omega_{\rm DM}$ as they both originated from one and the same axion field during the same QCD epoch 
as illustrated on Fig. \ref{phase_diagram}. 

To proceed with this task we first remind a simplified expression derived in our previous papers  \cite{Liang:2016tqc,Ge:2017ttc} where a number of effects have been neglected as mentioned in Section \ref{intro}.
In the simplified treatment  we characterized the system by a single crucial parameter $``c"$ which describes the disparity  between nuggets and antinuggets  and defined as follows\footnote{The definition of $``c"$ in present work  is   different comparing to previous studies \cite{Liang:2016tqc,Ge:2017ttc}, where $c\equiv B_{\rm antinuggets}/B_{\rm nuggets}$ was negative and inversely defined in contrast  with (\ref{eq:c_previous}).}: 
\begin{equation}
\label{eq:c_previous}
\Omega_{\rm dark}\simeq\left(\frac{1+c}{1-c}\right)\Omega_{\rm visible}, ~~~~c\equiv\frac{|B_{\rm nuggets}|}{|B_{\rm antinuggets}|}.
\end{equation}
The most important consequence of this model  is given by relation  (\ref{Omega}).
It   is expressed  now in terms of numerical parameter $c$  as  eq.  (\ref{eq:c_previous}) states. As we already stated this equation 
  represents a very  generic consequence of the entire framework. The parameter $0< c < 1$ is obviously positive and cannot be greater than 1 as eq. (\ref{eq:c_previous}) suggests. This is because the baryon charge hidden in the nuggets should be less than the baryon charge hidden in the antinuggets as the difference represents the baryon charge of the visible   matter in hadronic phase we presently observe. 
  
  As we already mentioned in Introduction in our previous  studies  \cite{Liang:2016tqc,Ge:2017ttc}  the relation  (\ref{eq:c_previous}) 
  is not accurate   as 
  we previously ignored a number of numerical factors. In particular, 
  we assumed that the energy per baryon charge  for hadronic   and CS phases is the same;   we ignored  the contribution of the axion domain wall surrounding the quark nugget in CS phase;  we neglected the contribution of  the conventional  propagating axions
produced by misalignment mechanism (or by decay of the topological defects)    by  assuming that  $\Omega_{\rm dark}$   is  saturated by the nuggets. 
All these simplifications allowed us to derive  a simple formula (\ref{eq:c_previous}) in terms of a single parameter $c$, which is very important for qualitative, but not quantitative,  arguments.  
Now we want to generalize relation  (\ref{eq:c_previous}) by accounting  for all these (previously neglected) effects. The corresponding modifications obviously    do not  affect  the basic qualitative claim (\ref{Omega}), but may  change some  numerical factors, which is precisely the  main objective of the present studies.  

Traditionally, the axion is regarded as one of the leading candidates for DM,
 see e.g. recent reviews~\cite{Marsh:2015xka,Graham:2015ouw,Ringwald2016} and references therein. It is normally assumed that the initial coherent axion field radiates  the propagating axions through  the misalignment mechanism or as a result of decay of the topological defects.  
  
 However, 
these two distinct processes of axion's  field relaxation  will be always accompanied by another complimentary mechanism (related to  the so-called $N_{\rm DW}=1$ axion domain walls) as recently advocated in refs.  \cite{Liang:2016tqc,Ge:2017ttc}. 
 The corresponding scenario is overviewed in Sections \ref{intro} and \ref{sec:AQN}, and 
  shall not be elaborated  here. The only important elements to be mentioned now are:  a) this  scenario  assumes that 
 the  PQ symmetry breaking occurs  before or during inflation,  and b)  this scenario  leads to a very  generic consequence  (\ref{Omega})
 which holds irrespectively   to any parameters of the system such as $m_a$ or initial misalignment angle $\theta_0$.
 This generic prediction of this new proposal  should be  contrasted with  two  other well known and well studied  mechanisms of the axion field's  relaxation when the corresponding contribution to the DM density  is highly sensitive to 
 the axion parameters,  $  \Omega_{\rm dark}\sim \theta_0^2/m_a^{7/6}$.

\exclude{for certain range of axion decay constant $f_{a}$ (or axion mass $m_{a}$ correspondingly), the value of initial axion misalignment angle $\theta_{a,i}$ has to be fine-tuned very close to 0 or $\pi$, to make the axion fully form DM~\cite{Wantz:2009it,Marsh:2015xka}. Counting (anti)nuggets as part of DM will avoid this fine-tunning problem, and for the allowed parameter space of $m_{a}$ and $\theta_{a,i}$, there will always be solutions to our DM model. This will be discussed in details in section {\ref{sec:4.c}}. 
}

We now proceed with our definitions and notations  of  the AQN model which incorporates both contributions: the conventional axion production reviewed in \cite{Marsh:2015xka,Graham:2015ouw,Ringwald2016} and the nugget's contribution studied in \cite{Liang:2016tqc,Ge:2017ttc}.  
  As usual, we use the ratio of different components' densities to critical density today to mark their proportions
\begin{equation}
\label{eq:Omega}
\Omega_{i}=\rho_{i}/\rho_{\rm cr},~~~~i=b,\pm,a,
\end{equation}
where label $i$ stands for different species: $b$ for baryon; `$+$' for nugget and `$-$' for antinugget; $a$ for free axions. $\rho_{\rm cr}$ is the critical density of the Universe today. Then we have
\begin{equation}
\label{eq:ratio}
\Omega_{DM}:\Omega_{b}=(\Omega_{+}+\Omega_{-}+\Omega_{a}):\Omega_{b}\approx5:1,
\end{equation}
where $\Omega_{DM}$ is represented by two parts, the axion contribution   $\Omega_{DM} (\rm axion)\equiv \Omega_{a}$ and the nugget's portion $\Omega_{DM} (\rm nugget)\equiv 
(\Omega_{+}+\Omega_{-})$.

Next, to describe the difference between nugget and antinugget, we define the following parameters:
\begin{equation}
\label{eq:c_Omega}
c_{\Omega}\equiv\frac{\Omega_{+}}{\Omega_{-}}=\frac{N_{+}E_{+}}{N_{-}E_{-}}=\frac{N_{+}\epsilon_{+}B_{+}}{N_{-}\epsilon_{-}B_{-}}=\frac{N_{+}}{N_{-}}c_{\epsilon}\cdot c,
\end{equation}
where
\begin{equation}
\label{eq:c&c_epsilon}
c_{\epsilon}\equiv\frac{\epsilon_{+}}{\epsilon_{-}},\qquad
c\equiv\frac{B_{+}}{B_{-}}, \qquad 0< c<1, 
\end{equation}
and $N_{\pm}$ is the number density of (anti)nuggets in the Universe today; $E_{\pm}$, $B_{\pm}$ and $\epsilon_{\pm}=E_{\pm}/B_{\pm}$ are respectively the energy, baryon number and energy density per baryon charge for a single  nugget   or antinugget. In these relations both baryon charges $B_{+}$ and $B_{-}$ are defined   to be  positive numbers.  The definition of $c$  in (\ref{eq:c&c_epsilon}) coincides with the definition given in (\ref{eq:c_previous}).

As we already mentioned after eq.  (\ref{eq:c_previous}) we neglected a number of important numerical factors in our previous studies ~\cite{Liang:2016tqc,Ge:2017ttc}. We can now formalize these effects in a very precise way using our definitions   (\ref{eq:c_Omega}) and (\ref{eq:c&c_epsilon}). 
In particular,  $\epsilon_{\pm}$  is the energy per baryon charge in CS phase is not the same as in hadronic phase, i.e. $\epsilon_{\pm}\neq m_p/3$. Furthermore, 
   $E_{\pm}$  which was previously estimated   as $E_{+}=E_{-}=  B m_p$ now  includes the contribution from the surrounding axion domain wall, and obviously has much more complicated structure. 
   In addition, we previously ignored the contribution from free propagating axions  by assuming $\Omega_{a}=0$.
      All these new elements will be incorporated into our analysis  in section~\ref{sec:3.epsilon model}. 

To simplify our analysis we want to make a technical assumption that $N_+\simeq N_-$ in eq. (\ref{eq:c_Omega}).  It  does not affect any of our conclusions as we argue below.  The basic justification for this assumption is as follows: the initial distributions of the nuggets and antinuggets is the same. It is just their evolution in the background of the coherent axion $\cal{CP}$-odd field generates  the asymmetry between them as argued in \cite{Ge:2017ttc}. 
As  time evolves the nuggets and antinuggets assume different size distribution, but their number densities,  $N_+$ and  $N_-$ should stay the same\footnote{There is some loophole in this argument as smaller bubbles may completely collapse, while larger bubbles survive the evolution. As the size-distribution for $B_-$ is 
different from $B_+$ the corresponding number densities  for $N_+$ and  $N_-$ could be also different as a result of different rates for collapsed nuggets. We expect that this effect is quite small. In any event, this effect  can be easily accommodated by redefinition of  the effective $B^{\rm eff}_+$ and  $B^{\rm eff}_-$ accounting for the collapsed nuggets.}. This information is coded  in equations (\ref{eq:c_Omega}) and (\ref{eq:c&c_epsilon}) by the relation   $B_+\neq B_-$. 

 Substituting $N_+= N_-$ in eq. (\ref{eq:c_Omega}) we arrive to desired expression 
\begin{equation}
\label{eq:c_relation}
c_{\Omega}=c_{\epsilon}\cdot c.
\end{equation}
This equation has obvious physical meaning and essentially states that the relative contribution of the nuggets and antinuggets to $\Omega_{DM}$ 
is proportional to the corresponding baryon charges expressed in terms of parameter $\sim c$, and it is also proportional to the difference of  their 
energy densities parametrized by $c_{\epsilon}$.

Now we want to derive an equation, similar to  (\ref{eq:c_previous}) which accounts for a number of the effects which were previously ignored, including the differences between nuggets and antinuggets. With this purpose we express the 
baryon number conservation  in the following form
\begin{equation}
\label{eq:NN}
 B_{-}- B_{+} \simeq \frac{1}{N_\pm}\frac{3\rho_{b}}{m_{p}},
\end{equation}
where $m_{p}$ is the mass of a single baryon charge and can be approximated  by the proton mass. The coefficient  3 in eq. (\ref{eq:NN}) corresponds to    our normalization of the baryon charge in the present work. This normalization is consistent with our definition of   $\mu$ corresponding to  the quark (rather than  baryon) chemical potential.  Furthermore,  $B_{\pm}$ in all our formulae are positive numbers as mentioned above. Therefore, the    $B_{\pm}$ parameters count for the number of quarks in the system, rather then their baryon charges. 
 
With these comments in mind and using  eqs.($\ref{eq:Omega}$), ($\ref{eq:c_Omega}$) and ($\ref{eq:NN}$), we arrive to the following relations
\begin{equation}
\label{eq:secondlast}
\frac{\Omega_{+}}{\epsilon_{+}}(\frac{1}{c}-1)=\frac{\Omega_{-}}{\epsilon_{-}}(1-c)=\frac{3\Omega_{b}}{m_{p}}.
\end{equation}
 The coefficient  $c\in (0, 1)$ in these relations satisfies the same constraint as in our simplified treatment of the problem presented in (\ref{eq:c_previous}). Therefore, 
\begin{equation}
\label{eq:c_constraint}
0<c<1.
\end{equation}
Our next step is to rewrite  the equations  ($\ref{eq:ratio}$) and ($\ref{eq:c_Omega}$) in the following convenient form 
\begin{equation}
\label{Omega_pm}
\Omega_{+}=c_{\Omega}\cdot\frac{\Omega_{DM}-\Omega_{a}}{1+c_{\Omega}},~~~~\Omega_{-}=\frac{\Omega_{DM}-\Omega_{a}}{1+c_{\Omega}}.
\end{equation}
The last step to achieve our goal  is to use   eqs.($\ref{eq:secondlast}$) and  (\ref{eq:c_relation}) to arrive to the final expressions which will be used in  our numerical studies  in Section\ref{sec:4.c}: 
\begin{subequations}\label{eq:results}
	\begin{align}
	\frac{1-c}{1+c_{\epsilon}c}=\frac{3\epsilon_{-}}{m_{p}}\cdot\frac{\Omega_{b}}{\Omega_{DM}-\Omega_{a}},\label{eq:results1}\\
	c_{\epsilon}\cdot\frac{1-c}{1+c_{\epsilon}c}=\frac{3\epsilon_{+}}{m_{p}}\cdot\frac{\Omega_{b}}{\Omega_{DM}-\Omega_{a}}.\label{eq:results2}
	\end{align}
\end{subequations}
The equation (\ref{eq:results}) is a generalization of our previous simplified expression (\ref{eq:c_previous}).
It accounts for a number of numerical effects mentioned previously.  Formula  (\ref{eq:results}) obviously reduces to our previous expression (\ref{eq:c_previous}) in the limit 
when $\Omega_{a}=0$ and nuggets and antinuggets have the same energies, i.e., $\epsilon_{+}=\epsilon_{-}=m_{p}/3$ such that $c_{\epsilon}=1$.

While the numerical estimates for parameters  $3\epsilon_{\pm}/m_{p}$ and $c_{\epsilon}$  entering (\ref{eq:results}) will be discussed in details in next section \ref{sec:3.epsilon model},   the rest of this section is devoted to  a short overview of known  estimates of the parameter $\Omega_{a}$ which  also enters eq.(\ref{eq:results}).  

The corresponding computations   of  $\Omega_{a}$ have been carried out in a number of papers. 
 In what follows   we limit ourselves by reviewing    the estimates of  $\Omega_{a}$    resulted from the  misalignment mechanism  \cite{misalignment},
while  leaving out   the contribution related to the decay of the topological defects\footnote{\label{defects}There is a number of uncertainties and  remaining discrepancies in the corresponding estimates. We shall not comment on these subtleties by referring to the original   papers \cite{Chang:1998tb,Hiramatsu:2012gg,Kawasaki:2014sqa,Fleury:2015aca,Klaer:2017ond}. 
  According to the most recent computations presented in ref.\cite{Klaer:2017ond}, the axion contribution to $\Omega_{\rm DM}$ as a result of decay of the  topological objects can saturate the observed DM density today if the axion mass is  in the range $m_a=(2.62\pm0.34)10^{-5} {\rm eV}$, while the earlier estimates suggest that the saturation occurs at a larger axion mass. One should also emphasize that the computations  \cite{Chang:1998tb,Hiramatsu:2012gg,Kawasaki:2014sqa,Fleury:2015aca,Klaer:2017ond} have been performed with assumption that PQ symmetry was broken after inflation.}.
  We emphasize that we exclude the  corresponding contributions related to the  topological defects not because they are unimportant. Rather, we omit them because their role  is largely  unknown under present circumstances when the PQ symmetry is broken before/during inflation\footnote{It was  explained  in \cite{Liang:2016tqc}  in great details that the so-called $N_{\rm DW}=1$ domain walls must be present in the system even when PQ symmetry was broken before/during inflation, see item 5 in section III of \cite{Liang:2016tqc}. It should be contrasted with conventional studies
   on the role of the topological defects when the PQ symmetry is  assumed to  be broken after the inflation, see also  footnote \ref{defects} with related comment.}.  In addition, even in a different scenario when  the PQ symmetry broken after inflation, the question  whether it saturates the observed dark matter density  remains controversial as mentioned in footnote \ref{defects}.  Thus,
  we leave out this contribution to simplify our notations and our analysis as the focus of the present work is the nugget's contribution  to $\Omega_{\rm DM}$
  rather than the direct axion production represented by $\Omega_{a}$. In other words,  
 $\Omega_{a}$ contribution  is kept in our formulae for normalization purposes to illustrate   the significance (or insignificance) of the nugget's contribution to $\Omega_{\rm DM}$ as a function of parameters. The contribution to the   $\Omega_{a}$ due to the decay of the topological objects can be always incorporated into our formulae once the  uncertainties  of this contribution  are better understood.

 For the vacuum misalignment production of free propagating axions, we adopt the general formula as presented in ref.~\cite{Marsh:2015xka}
\begin{equation}
\label{eq:Omega_a_full}
\begin{aligned}
\Omega_{a}h^2&\approx2\times10^4\left(\frac{f_{a}}{10^{16}{\rm GeV}}\right)^{7/6}\left(\theta_{a,i}^2+\frac{H_{I}^2}{(2\pi f_{a})^2}\right)\\
&\times F_{\rm anh}\left(\sqrt{\theta_{a,i}^2+\frac{H_{I}^2}{(2\pi f_{a})^2}}\right),
\end{aligned}
\end{equation}
with 
\begin{equation}
F_{\rm anh}(x)=\left[\ln{\left(\frac{e}{1-x^2/\pi^2}\right)}\right]^{7/6},
\end{equation}
where $H_{I}$ is the inflationary Hubble scale and $F_{\rm anh}(x)$ is the correction factor due to the anharmonic cosine part in axion potential~\cite{Marsh:2015xka,Visinelli:2009zm}. The parameter $\theta_{a,i}$ 
in (\ref{eq:Omega_a_full}) is the initial misalignment angle and $H_{I}/(2\pi f_{a})$ is the backreaction contribution to this homogeneous field displacement due to the isocurvature perturbations. The parameter $f_{a}$
and  the axion mass $m_{a}$ are not independent parameters as their product is fixed by the topological susceptibility of QCD, $\chi=f_{a}^{2}m_{a}^{2}$. Using the recent value $\chi=0.0216$ fm$^{-4}$ at zero temperature~\cite{Borsanyi:2016ksw}, we have
\begin{equation}\label{eq:mafa}
m_{a}\simeq5.7\times10^{-4}{\rm eV}\left(\frac{10^{10}{\rm GeV}}{f_{a}}\right).
\end{equation}
This completes our short overview of $\Omega_{a}$ contribution entering our basic formulae (\ref{eq:results}).

The  next  section \ref{sec:3.epsilon model} is mainly  devoted to the estimates of  $3\epsilon_{\pm}/m_{p}$ and $c_{\epsilon}$  which, along with  $\Omega_{a}$ also enter our basic equation (\ref{eq:results}). Then in section \ref{sec:4.c}, using the results of sections \ref{sec:2.density} and   \ref{sec:3.epsilon model}, we do numerical analysis to  study 
the allowed window and constraints related to the phenomenological parameters of the AQN dark matter model.

 We conclude this section with the following generic comment. The nugget's contribution given by $\Omega_{\pm}$ and the direct axion production represented by $\Omega_{a}$ always accompany each other  during   relaxation of the  dynamical  axion field   to zero at the QCD epoch. These contributions to $\Omega_{\rm DM}$ represent complementary mechanisms  and cannot be formally separated (e.g. by variation
 of  a free parameter  of the system such as $f_a$) as the closed $N_{\rm DW}=1$ domain wall bubbles, which are responsible for the nugget's formation,  can be produced irrespectively whether the PQ scale is above or below the inflationary scale $H_{I}$,  as argued in  \cite{Liang:2016tqc}.

\section{Internal structure of the nuggets}
\label{sec:3.epsilon model}
The main goal of this section is to estimate   the energy per baryon charge of a stabilized  quark nugget at zero temperature.  We consider two drastically different  models to accomplish this task. The main reason to consider two different models (constructed on fundamentally  different principles)  is to test the sensitivity (or non-sensitivity) to different phenomenological parameters effectively describing the strongly coupled QCD. For simplicity and  without loss of generality, we  assume that  the CS phase  assumes the simplest  possible structure in the form of the  
 colour flavour locked (CFL) phase without any additional complications such as  possible meson condensation.

The first model largely follows the original work \cite{Zhitnitsky:2002qa}. 
However, the difference with the previous  analysis  is that the  first paper on the subject  \cite{Zhitnitsky:2002qa}  was mostly dealing with   fundamental and basic questions on  principle possibility to stabilize the nuggets by the axion domain walls. 
The goal of the present studies is quite  different as we want to  produce some quantitative results 
on the parameters entering the basic equation  (\ref{eq:results}).

 The model  \cite{Zhitnitsky:2002qa}
 considers the equilibration between  the Fermi pressure,  the domain wall surface tension, the  ``bag constant'' pressure $\sim E_B$, and, finally,  the quark-quark interaction  related  to the CS gap. The energy of a stabilized nugget can be represented in the following form 
\be
\label{eq:3.Model 1 E}
E^{(1)}
&=&4\pi\sigma_{\rm eff}R^2+\frac{g\mu^4}{6\pi}R^3
-\frac{3\Delta^2\mu^2}{\pi^2}V  \\
&+&E_B\theta(\mu-\mu_1)\left(1-\frac{\mu_1^2}{\mu^2}\right)V, ~~~~ V\equiv \frac{4\pi}{3}R^3, \nonumber
\ee
while  the nugget's baryon number can be estimated as follows 
\begin{equation}
\label{eq:3.Model 1 B}
B
=gV\int_0^\mu\frac{d^3p}{(2\pi)^3}
=\frac{2g}{9\pi}\mu^3R^3, 
\end{equation}
where our normalization corresponds to $B=1$ per single quark degree of freedom in order to remain    consistent with notations  of Sec. \ref{sec:2.density}.

Few comments are in order. First of all,   the ``squeezer'' parameter $\mu_1$ is estimated to be 330 MeV, the degeneracy factor for CFL phase is estimated as $g\simeq 2N_cN_f\simeq 18$. Parameters $\mu$ and $R$ entering (\ref{eq:3.Model 1 E}) and (\ref{eq:3.Model 1 B}) are the chemical potential and radius of the nugget, respectively, while the bag constant $ E_B\simeq(150{\rm ~MeV})^4$ and the CS gap $\Delta\simeq100 {\rm ~MeV}$ assume their commonly accepted magnitudes, see original work \cite{Zhitnitsky:2002qa} with 
some arguments supporting these numerical values. 

The domain wall tension $\sigma_{\rm eff}$ entering (\ref{eq:3.Model 1 E}) requires some additional comments. First of all,  
  the effective domain wall tension $\sigma_{\rm eff}$ should not be confused with  the conventional surface tension $\sigma\simeq 8f_a^2m_a$
  which   normally enters the computations  \cite{Chang:1998tb,Hiramatsu:2012gg,Kawasaki:2014sqa,Fleury:2015aca,Klaer:2017ond} of the axion production due to the decay of the topological defects. 
  
  There are two main reasons for this important difference. First of all, the axion domain wall   solution in our case interpolates between topologically distinct  vacuum states in hadronic and CS phases, in contrast with conventional axion domain wall which interpolates between topologically distinct  hadronic vacuum states.  The chiral condensate may or may not be formed in CS phase. It strongly affects 
  the topological susceptibility in CS phase which could be much smaller than in the conventional hadronic phase.
  The well known manifestation of this difference is the expected smallness of the  $\eta'$ mass in CS phase in comparison with the hadronic phase.  
One should emphasize the $2\pi$ periodicity of the axion ($\theta$) and the Nambu-Goldstone fields ($\eta'$) still hold in presence of the chemical potential $\mu$ in dense matter CS phases \cite{Son:2000fh}. Therefore, the topological reason for mere existence of the axion domain wall still persists, while the numerical value of the tension $\sigma_{\rm eff}$ will deviate from its conventional expression $\sigma$ computed in hadronic phase.

  The second  reason for strong deviation of the $\sigma_{\rm eff}$ from conventional expression for $\sigma$ is that formula $\sigma\simeq 8f_a^2m_a$
   was derived assuming the thin-wall approximation when the domain wall is assumed to be almost flat, i.e. a typical curvature of the domain wall structure is much smaller than its width. This approximation is obviously badly violated 
   because  the axion domain wall   width is of order $m_a^{-1}$, 
   while typical curvature is   comparable with the width of the domain wall as these two parameters are related in our framework,   $R\sim m_a^{-1}$. The physical consequence of this relation is that the axion field 
   strongly overlaps within the nugget's volume. This effect is expected to   drastically  reduce  the domain wall tension\footnote{The corresponding large modifications can be understood from a simple model when the domain wall is bended allowing a strong overlap between opposite sides of the wall. The effective domain wall tension   obviously receives the modifications as a result of this bending  geometry when the axion field configuration deviates from a simple well-known 1-dimensional solution.}.

       To account for these complicated QCD effects we define  $\sigma_{\rm eff}\equiv\kappa\cdot\sigma$, with an unknown phenomenological parameter $0<\kappa<1$ which accounts for the physics mentioned above. In particular, the violation of the thin-wall approximation was modelled  in  \cite{Liang:2016tqc}
       by introducing a suppression factor   $ \exp({-R_0/R_{\rm form}})$.  
  The corresponding suppression could be quite strong and can be   as small as $10^{-5}$ assuming a typical  formation radius $R_{\rm form}\sim0.1R_0$ as studied in  \cite{Liang:2016tqc}. In what follows we treat $\kappa$ as a free phenomenological parameter.

  Our goal now is to minimize the expression (\ref{eq:3.Model 1 E}). To achieve this goal we introduce  
  two dimensionless variable $x$ and $\sigma_0$ as
\begin{equation}
\label{eq:3.Model 1 x sigma}
\begin{aligned}
x&\equiv R\frac{E_B^{1/4}}{B^{1/3}}
=\frac{E_B^{1/4}}{\mu}\left(\frac{8\pi}{2g}\right)^{\frac{1}{3}},  \\
\sigma_0&\equiv\frac{\sigma_{\rm eff}}{B^{1/3}E_B^{3/4}}
=\frac{8\chi}{E_B^{3/4}}\frac{1}{\kappa^{-1}B^{1/3}m_a},
\end{aligned}
\end{equation}
where we express $\sigma$ in terms of topological susceptibility  $\chi=f_a^2m_a^2=0.02{\rm fm}^{-4}$. Then, the energy per quark $\epsilon\equiv E/B$  can be expressed as 
\begin{equation}
\label{eq:3.Model 1 epsilon}
\begin{aligned}
\epsilon_{\rm tot}^{(1)}(x)
&\equiv\frac{E^{(1)}}{B}=\epsilon_{\rm DW}+\epsilon_{\rm QCD}^{(1)}(x),  \\
\epsilon_{\rm DW}(x)
&=E_B^{1/4}4\pi\sigma_0x^2
=\frac{32\pi\chi}{E_B^{1/2}}\frac{x^2}{\kappa^{-1}B^{1/3}m_a},  \\
\epsilon_{\rm QCD}^{(1)}(x)
&=\frac{3}{4}\left(\frac{9\pi}{2g}\right)^{\frac{1}{3}}\frac{E_B^{1/4}}{x}
-\frac{18\Delta^2}{gE_B^{1/4}}\left(\frac{2g}{9\pi}\right)^{\frac{1}{3}}x  \\
&\quad+\frac{4\pi}{3}E_B^{\frac{1}{4}}\theta(x_1-x)\left(1-\frac{x^2}{x_1^2}\right)x^3, \\
\end{aligned}
\end{equation}
The equilibrium point can be found following the condition
\begin{equation}
\label{eq:3.Model 1 equilibrium}
0=\left.\frac{\partial \epsilon_{\rm tot}}{\partial x}\right|_{x=x_{\rm eq.}}.
\end{equation}
The solution can be well approximated from numerical computation as
\begin{subequations}
\label{eq:3.Model 1 epsilon_final}
\begin{equation}
\label{eq:3.Model 1 epsilon_final_a}
\epsilon_{\rm tot}^{(1)} 
\simeq-0.57m_\pi+\frac{3.51m_\pi}{(\kappa^{-1}B^{1/3}\frac{m_a}{m_\pi})^{0.310}},
\end{equation}
\begin{equation}
\label{eq:3.Model 1 epsilon_final_b}
{\rm within\qquad} 0.2\lesssim\kappa^{-1}B_\pm^{1/3}m_a/m_\pi\lesssim0.95.
\end{equation}
\end{subequations}
This solution has accuracy up to  $0.6\%$ comparing to the exact numerical solution within the range \eqref{eq:3.Model 1 epsilon_final_b}, see Appendix \ref{appendix:A.model comparison} with more technical details. This energy per baryon charge will be used in next section for our numerical analysis of this AQN dark matter model to test  its consistency with all available constraints.

Our second model to be considered has very different building principle, and essentially is based on the ideas of   old constituent quark model being applied to the dense matter  systems  \cite{Kojo:2014rca}.  
 This model will serve  as a complementary tool which allows us to test  the sensitivity  of our results to different phenomenological parameters effectively describing the strongly coupled dense QCD.  
 
 We assume the following  form for the energy of the nuggets 
\begin{equation}
\label{eq:3.Model 2 E}
E^{(2)}=BM_q+4\pi\sigma_{\rm eff}R^2.
\end{equation}
Here $M_q$ is the effective constituent quark's mass  in CS phase in the bulk of  the nugget. The  $M_q$  should have a typical QCD scale  and  serves as a parameter of the model. The energy per quark degree of freedom is therefore
\begin{equation}
\label{eq:3.Model 2 epsilon}
\begin{aligned}
\epsilon_{\rm tot}^{(2)} 
&=M_q+\frac{32\pi\chi}{\mu^2}\left(\frac{9\pi}{2g}\right)^{\frac{2}{3}}
\frac{1}{\kappa^{-1}B^{1/3}m_a}  \\
&\simeq M_q+\left(\frac{376{\rm MeV}}{\mu}\right)^2
\frac{m_\pi}{\kappa^{-1}B^{1/3}\frac{m_a}{m_\pi}},
\end{aligned}
\end{equation}
where we substitute $\frac{R}{B^{1/3}}=\frac{1}{\mu}\left(\frac{9\pi}{2g}\right)^{\frac{1}{3}}$ from Eq. \eqref{eq:3.Model 1 B}. We should note that $M_q$ and $\mu$ are not completely free parameters according to various phenomenological models for dense phases. To be consistent with previous studies (applied to the neutron star physics where this model was used) we adopt the following numerical values for relevant parameters \cite{Kojo:2014rca}:
  \begin{equation}
\label{eq:3.Model 2 M mu}
(M_q,~\mu)\simeq(200,~400){\rm ~MeV},~{\rm and~}(160,~500){\rm ~MeV}.
\end{equation}
In spite of the fact that the Model-2 has a fundamentally different building principle,  we observed  that these choices (\ref{eq:3.Model 2 M mu}) for $(M_q,~\mu)$ parameters produce the  results which numerically very close to the results \eqref{eq:3.Model 1 epsilon_final}  obtained for  Model-1, see  Appendix \ref{appendix:A.model comparison}
with detail analysis. 

We conclude this ``technical section" with the following comment. Both models discussed in this section  and represented by eqs.\eqref{eq:3.Model 1 epsilon_final} and \eqref{eq:3.Model 2 epsilon} are based on  fundamentally different principles. Nevertheless,  both models lead to similar results for relevant parameters, and furthermore, demonstrate that the energy density $\epsilon$ in both models depends on a  single dimensionless parameter   $\kappa^{-1}B^{1/3}m_a/m_\pi$, which is clearly a highly nontrivial feature. A more detailed discussions   of these two models can be found  in Appendix \ref{appendix:A.model comparison}.

\section{The AQN    model confronting the observations}\label{sec:4.c}

The main purpose of this section is to analyze all possible constraints on the parameters of  the AQN  model.
In previous sections we introduced  a number of phenomenological parameters describing this   model. Our goal here is to analyze   the allowed region in parametrical space where these parameters may vary but remain  consistent  with all known observations/experiments. 
We use the  result (\ref{eq:3.Model 1 epsilon_final}) or \eqref{eq:3.Model 2 epsilon} from section~\ref{sec:3.epsilon model} to fit the   parameters $\epsilon_{\pm}$. We use (\ref{eq:Omega_a_full}) from section~\ref{sec:2.density} to fit   the density of the relic  dark matter  propagating axions into our AQN model.   Finally, we implement constraints on  physically observable parameters $B$,   $\theta_{a,i}$, $m_{a}$, $H_{I}$ from large number of independent  experiments and observations to analyze  the  allowed window for most important  parameter of the AQN model represented by the coefficient $``c"$. This key parameter is  defined by eq. (\ref{eq:c&c_epsilon}) and describes the disparity between nuggets and antinuggets.  As we argued in  \cite{Ge:2017ttc} this coefficient $c$ must be order of one as a result of interaction with    coherent $\cal{CP}$ odd axion field. This coefficient, in principle, is calculable from the first principles
along with other parameters of the model as all fields, coupling constants and interactions are represented by the SM physics accompanied by the axion field $\theta(x)$ with a single additional fundamental parameter $f_a$. However, such computations presently are not feasible as even the phase diagram shown on Fig.\ref{phase_diagram} at $\theta\neq 0$ is not yet understood.
 
 We represent our  numerical results  in subsections~\ref{subsec:plots} and \ref{subsec:tuning}. However,  first of all, in next subsection~\ref{subsec:constraints}  we overview the known  constraints on relevant parameters  of the AQN model. 

\subsection{Constraints on parameters of the AQN model}\label{subsec:constraints}
We start with  eq. (\ref{eq:c_constraint}) which defines coefficient  $c$. This should not be considered as a constraint on $c$ as it essentially represents 
our convention that we define the visible matter as the baryons with positive baryon charge. Therefore, the absolute value of the baryon charge hidden in the antinuggets $B_-$ must be greater than the baryon charge $B_+$ hidden in the nuggets (as a result of global conservation of the baryon charge which is assumed to be zero at all times).  This leads to  the formal relation  $c< 1$ which reflects our convention. The parameter $c$ is obviously a positively defined  parameter which is  explicitly represented by  eq. (\ref{eq:c&c_epsilon}).
 
Another constraint $0.2\lesssim\kappa^{-1}B^{1/3}m_{a}/m_\pi\lesssim0.95$ follows  from (\ref{eq:3.Model 1 epsilon_final_b}). 
This constraint is related to our studies  of the stability of the nuggets in CS phase,  see section~\ref{sec:3.epsilon model} with  more technical details. 

The next item we want to discuss is the known constraints on the baryon charge $B$ of the nuggets (or antinuggets) represented by (\ref{B-range}).
We already made few comments about    the constraints related to the galactic observations, Apollo data, and  ancient Mica analysis as mentioned   in section \ref{sec:AQN}, footnote \ref{B-constraint}  and reviewed   in \cite{Jacobs:2014yca,Lawson:2013bya}. Now we want to make few comments   regarding  other constraints related to different observations and analysis. 

 First of all we want to mention the constraints related to limits from the total geothermal energy budget of the Earth \cite{Gorham:2012hy}
 which is consistent with (\ref{B-range}).
It has also been suggested that the \textsc{anita} 
experiment may be sensitive to the radio band 
thermal emission generated by these objects as they pass through the 
antarctic ice \cite{Gorham:2012hy}. These experiments may thus be 
capable of adding direct detection capability to the indirect evidence 
mentioned previously  in section \ref{sec:AQN}. 

It has been also suggested recently \cite{Gorham:2015rfa} that  the interactions of  the antinuggets  with normal matter in the Earth and Sun will lead to annihilation and an associated neutrino flux. 
Furthermore, it has been claimed  \cite{Gorham:2015rfa} that the antiquark nuggets cannot account for more than 20$\%$ of the dark matter flux based on constraints for the neutrino flux in 20-50 MeV range where sensitivity of the underground neutrino detectors such as SuperK have their highest signal-to-noise ratio.  However, the claim  \cite{Gorham:2015rfa} was based on assumption that 
the annihilation of visible baryons with antiquark nuggets  generate  the neutrino spectrum    similar to conventional   baryon- antibaryon  annihilation  spectrum when the  large number of produced  pions   eventually decay to  muons 
and consequently to highly energetic neutrinos  in the 20-50 MeV energy range.  
 This claim has been dismissed in  ref. \cite{Lawson:2015cla} by emphasizing that antinuggets cannot be treated as an usual antimatter in conventional hadronic phase as the quarks in nuggets (and antiquarks in antinuggets) belong to CS phase rather than to the hadronic phase we are familiar with.
 
 One could also expect a  set of strong constraints on the antinuggets by considering the radio observations or strong 511 keV  line emission from nearby galaxies
 if one assumes that the antinuggets  (almost) saturate the corresponding emissions in our Milky Way. Indeed, similar consideration in other models essentially rules out the   dark matter explanation of 511 keV line observed  in our galaxy. However, the corresponding analysis carried out in   \cite{Lawson:2015xsq} for the radio radiation and in \cite{Lawson:2016mpu} for 511 keV line emission from  nearby galaxies does not lead to any new constraints, in huge contrast with conventional WIMPs models which typically predict  the radiation  from nearby galaxies  exceeding the observed values. We refer to
 the original papers  \cite{Lawson:2015xsq,Lawson:2016mpu} for the details and references. 
 
 Our last comment related to constraints on $B$ as given by (\ref{B-range}) is related to the recent arguments  \cite{Zhitnitsky:2017rop} advocating that 
 that  the extreme ultra violet (EUV) and soft x-ray emission from the solar corona might be  due to the antiquark nuggets hitting  the Sun from outer space. 
If one identifies   the observed ``nano-flares"   with annihilation events of the antinuggets in the solar corona one can 
infer the  nugget's baryon charge distribution  from the (measured)
 energy distribution of the ``nano-flares"   in solar corona,    see also some related comments in concluding section \ref{sec:5.conclusion}.    This identification leads  to an independent  estimate  \cite{Zhitnitsky:2017rop} on possible values for $B$ which is  perfectly consistent with the previous studies, which we 
 express as 
\begin{equation}\label{eq:B_constraint}
10^{23}\lesssim B\lesssim10^{28}.
\end{equation}
 In our numerical analysis     which follows we treat the window (\ref{eq:B_constraint}) as the solid constraint  of  the allowed magnitude of  the nugget's baryon charge.

We next consider the  classical  window\footnote{For the main purposes  of this paper, we will only consider ``the classical axion window'', where the initial misalignment angle $\theta_{a,i}$ is not fine-tuned. Note that while the upper bound is a very solid constraint as it is given by stellar physics (e.g. see review \cite{Raffelt:2006cw}), the lower bound on the axion mass in Eq. \eqref{eq:ma_costraint} should be treated as an order of magnitude estimate provided that $\theta_{a,i}$  is not fine-tuned. 
If the fine tuning is allowed, the  $\theta_{a,i}$ may assume arbitrarily small value, in which case the corresponding lower bound on $m_a$ is shifted. The only exclusion interval  in this case $6\times10^{-13}{\rm eV}<m_a<2\times10^{-11}{\rm eV}$ is obtained from black hole superradiance effects \cite{Arvanitaki:2014wva}.}
 for axion mass, see  recent review paper ~\cite{Graham:2015ouw}:
\begin{equation}\label{eq:ma_costraint} 
10^{-6}{\rm eV}~\lesssim m_{a}\lesssim10^{-2}~{\rm eV}.
\end{equation}
By using the relation (\ref{eq:mafa}) this window for $m_a$ can be expressed  in terms of the corresponding  classical window for $f_a$: 
\begin{equation}\label{eq:fa_costraint}
5.7\times10^{8}~{\rm GeV}\lesssim f_{a}\lesssim5.7\times10^{12}~{\rm GeV}.
\end{equation}
One should  emphasize  that the constraint (\ref{eq:ma_costraint}) or equivalently (\ref{eq:fa_costraint}) is commonly  accepted 
axion window, and it is not, by any means,  originated from our analysis of the AQN model. Nevertheless, all our constraints depend on $m_a$ as it explicitly enters 
the equation (\ref{eq:3.Model 1 epsilon_final_b}).  

From these discussions it is clear that the axion mass $m_a$ plays a dual role in our analysis because it enters the  formulae related to the physics of the nuggets
as  equation (\ref{eq:3.Model 1 epsilon_final_b}) states. It also enters  the expression  (\ref{eq:Omega_a_full}) for $\Omega_a$.   It unambiguously implies that   the remaining portion of the dark matter represented by the nugget's contribution   (\ref{Omega_pm})  becomes also (implicitly) highly  sensitive  to $m_a$ through dependence of the axion portion of the dark matter represented by $\Omega_a$.

As we argue below,  for a  reasonable values of $\kappa$ in range  $10^{-4}\lesssim\kappa\lesssim10^{-2}$, the constraints 
 (\ref{eq:3.Model 1 epsilon_final_b}),  (\ref{eq:B_constraint}) and  (\ref{eq:ma_costraint}) become  mutually compatible  which we consider as a highly nontrivial consistency check as all the parameters entering these  relations  have been constrained by very different  physics related to independent observations, experiments and analysis.

The  next  constraint to consider is related to 
analysis of  the inflationary scale $H_I$ and the related constraints on the tensor-to-scalar ratio and the isocurvature perturbations. 
The basic assumption of this work is that PQ symmetry breaking occurs before/during inflation, in which case  
\begin{equation}\label{eq:PQ_constraint}
f_{a}>H_{I}/2\pi, 
\end{equation}
see e.g. \cite{Marsh:2015xka} for review. This assumption plays a crucial role in our analysis \cite{Liang:2016tqc,Ge:2017ttc} because the $\cal{CP}$ odd axion field
must be coherent on enormous scale of the entire Universe to separate  the baryon charges   on these gigantic  scales with the same sign of $\theta$. Precisely this coherent axion field generates the disparity between the nuggets and antinuggets which eventually leads to the generic and fundamental prediction (\ref{Omega}) of this entire framework. 

It is known that the Inflationary Hubble scale is tied to the value of the tensor-to-scalar ratio $r_T$  which measures $H_I$. Assuming a simplest single field  inflationary model, the non-observation of  the tensor modes $(r_T<0.12)$ imposes the  upper limit for the inflation scale, see \cite{Marsh:2015xka,Ade:2015tva}:
\begin{equation}\label{eq:HI_constraint}
H_{I}/2\pi\lesssim1.4\times 10^{13}~{\rm GeV}.
\end{equation}
Important comment here that (\ref{eq:HI_constraint}) is highly model-dependent  result and varies from one inflationary model to another.
It is presented here exclusively for illustrative purposes to provide some orientation with the relevant scales of the problem. 

 The isocurvature perturbations related to the axion field provides another independent constraint on $H_I$. We recall that the amplitude for the isocurvature power spectrum is determined by the following expression, see the  original papers  \cite{Linde:1985yf,Seckel:1985tj,Hertzberg:2008wr,Hamann:2009yf,Kobayashi:2013nva}  and review  article\cite{Marsh:2015xka}:
  \be
 \label{isocurvature}
 A_I=\left(\frac{\Omega_a}{\Omega_{\rm DM}}\right)^2\cdot \left(\frac{H_I}{\pi \phi_i}\right)^2, ~~~ \phi_i\equiv f_a \theta_{a,i}.
 \ee
 The corresponding isocurvature amplitude is strongly constrained by CMB analysis, $A_I/A_s < 0.038$, where $A_s$ is conventional amplitude for the scalar power. It is normally assumed that the non-observation of the isocurvature perturbation provides  a strong constraint on  axion properties in scenario where the PQ symmetry broken before/during inflation. 
 Our original comment here is that   the axion contribution represented by $\Omega_a$ in eq. (\ref{isocurvature}) to the dark matter density $\Omega_{\rm DM}$ could be numerically quite small in the AQN model as the nuggets in most cases  play  the dominant role  by saturating  the  dark matter density.  Such a scenario  drastically alleviates some severe constraints on parameters in conventional analysis where one normally assumes that the  axions saturate the dark matter density.

 We conclude this subsection with the following remark. 
 The conventional analysis on the relation between dark matter axions, inflationary scale, isocurvature perturbations very often assumes that the axions saturate the dark matter density. It should be contrasted with our AQN model where the axions themselves with the same $f_a$
 may contribute very little to $\Omega_{\rm DM}$ as  the dominant contribution may come form the nuggets which always 
 satisfy the relation  $\Omega_{\rm DM}\sim \Omega_{\rm visible}$ according to (\ref{Omega}) irrespectively to $f_a$ or initial misalignment angle $\theta_{a,i}$. It may alleviate some severe constraints on the parameters (such as $H_I, f_a, r_T, \theta_{a,i}$) which other  models normally face.

\subsection{Numerical plots}\label{subsec:plots}
The  goal of this subsection is to analyze the dependence of the internal (with respect to the AQN model) parameter $``c"$ from  external parameters of the system such as $B, \theta_{a,i}, m_{a}, H_{I}$
which are well defined observables irrespectively to the specific features of the AQN model. As the parameter $``c"$ cannot be negative or  larger than one, the corresponding plots provide us with information on the typical values of the external parameters $B, \theta_{a,i}, m_{a}, H_{I}$ when the AQN  model is self-consistent with all presently available  constraints. 
\begin{figure*}
	\centering
	\captionsetup{justification=raggedright}
	\begin{subfigure}[t]{.497\textwidth}
		\includegraphics[width=\textwidth]{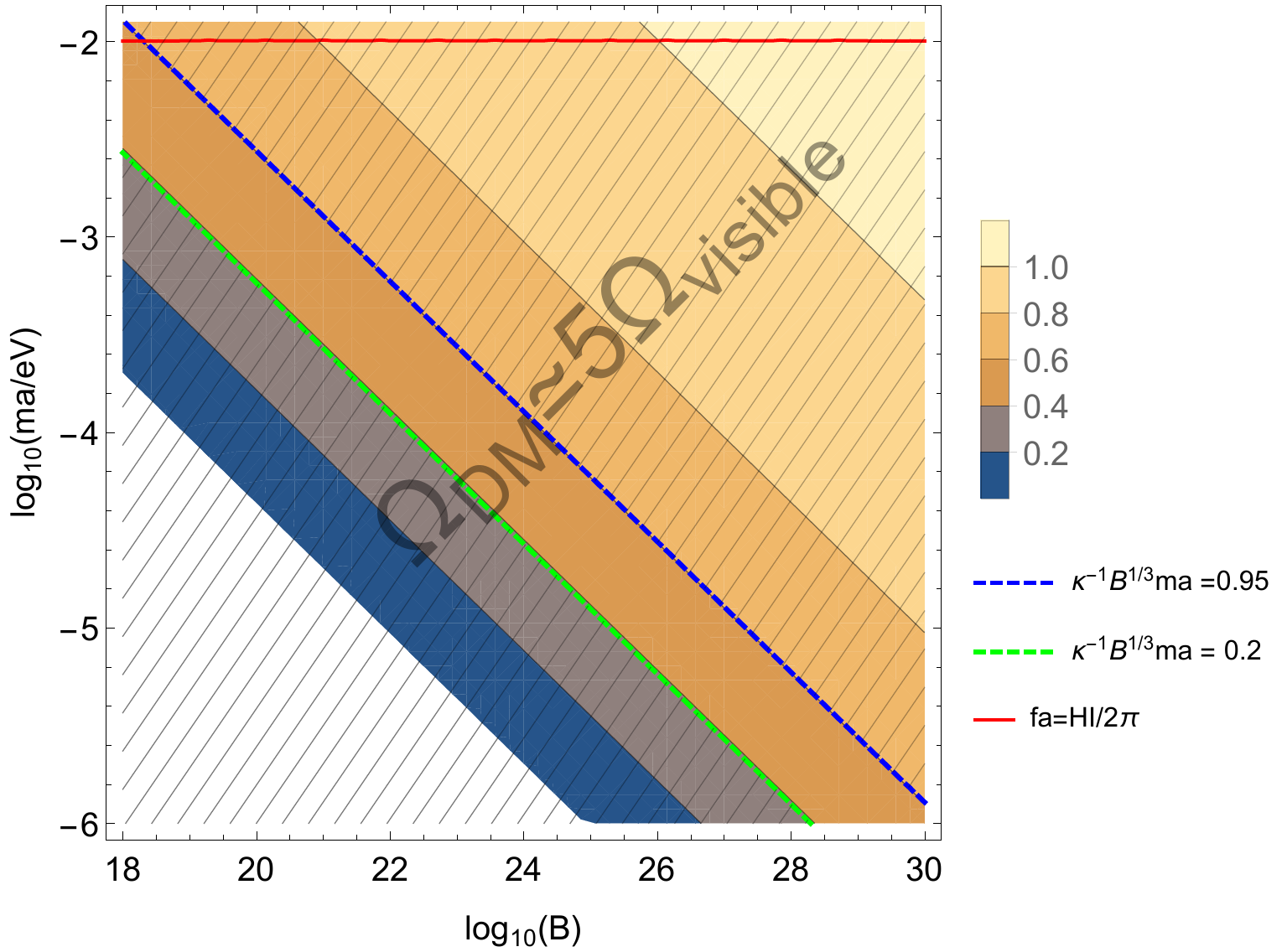}
		\caption{$H_{I}/2\pi=5.7\times10^{8}~{\rm GeV}, \theta_{a,i}=10^{-3}$}\label{c_HI=2pi57_ntheta=3}
	\end{subfigure}
	\begin{subfigure}[t]{.497\textwidth}
		\includegraphics[width=\textwidth]{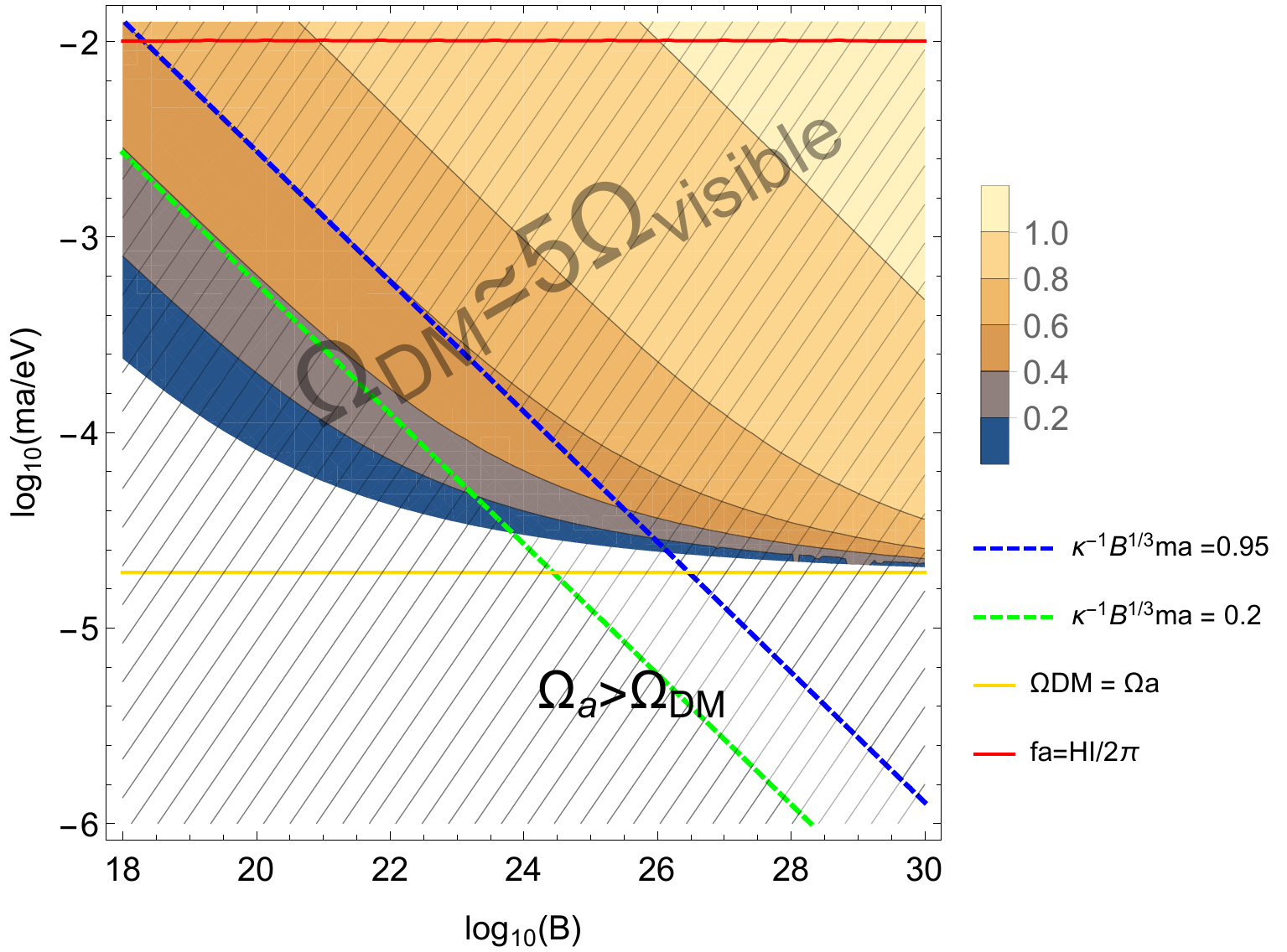}
		\caption{$H_{I}/2\pi=5.7\times10^{8}~{\rm GeV}, \theta_{a,i}=10^{0}$}\label{c_HI=2pi57_ntheta=0}
	\end{subfigure}
	\caption{Contour plots of $c$ as a function of $m_{a}$ and $B$ for $\theta_{a,i}=10^{-3}$ and $\theta_{a,i}=10^0$ respectively with the fixed values $H_{I}/2\pi=5.7\times10^8$ GeV and $\kappa=10^{-4}$. Different colours (shown on the right) correspond to the different values of the parameter $c\sim 1$. Precisely this parameter $c$ essentially determines the relation between dark and visible matters according to approximate relation (\ref{eq:c_previous}).  Here $B$ corresponds to  the baryon number of an antinugget (same for other figures in this subsection).}
	\label{fig:plotc1}
\end{figure*}

We start our analysis by plotting on  Fig.~\ref{c_HI=2pi57_ntheta=3}  the  parameter  $``c"$ as a function of $m_{a}$ and $B$, where we fix specific values for parameters $\kappa=10^{-4}$ and $H_I/2\pi=5.7\times10^8$ GeV and  $\theta_{a,i}=10^{-3}$  to simplify the arguments and analysis. We also plotted the $f_a=\frac{H_I}{2\pi}$ by red solid line to
localize  the physical parametric space and remove unphysical (within the AQN model) solutions. We also plotted (by green and blue dashed lines) the region in parametrical space where the condition (\ref{eq:3.Model 1 epsilon_final_b}) is satisfied and our computations in CS phase are justified.  
For this specific choice of the parameters one can  explicitly  see that parameter $c$ is constrained in a parallelogram with the range $0.4\lesssim c\lesssim0.6$. This region of the parametrical space satisfies all internal and external constraints listed in previous section. 

From the same plot one can   also
identify  the allowed region of the nugget's baryon charge $B$ for a given axion mass $m_a$. 
One should emphasize that the dark matter density on Fig.~\ref{c_HI=2pi57_ntheta=3}  for $\theta_{a,i}=10^{-3}$ is entire saturated by the nuggets as the direct axion production is strongly suppressed by small initial misalignment angle $\theta_{a,i}=10^{-3}$. To see the role of the direct axion production one can   choose $\theta_{a,i}=10^{0}$   as shown on Fig.~\ref{c_HI=2pi57_ntheta=0}.   In this case the direct axion production 
saturates the dark matter density $\Omega_{\rm DM}=\Omega_a$ at small axion  masses $m_a\simeq 10^{-5}$ eV, as shown by solid yellow line.   When $10^{-3}< \theta_{a,i}< 1$ varies  between these two values the corresponding allowed region  for the nuggets's parametrical space  $(B, m_a)$ will be modified accordingly as  the allowed region for the nuggets obviously shrinks when the direct axion production starts to play an essential role.  
\begin{figure*}
	\centering
	\captionsetup{justification=raggedright}
	\begin{subfigure}[t]{.497\textwidth}
		\includegraphics[width=\textwidth]{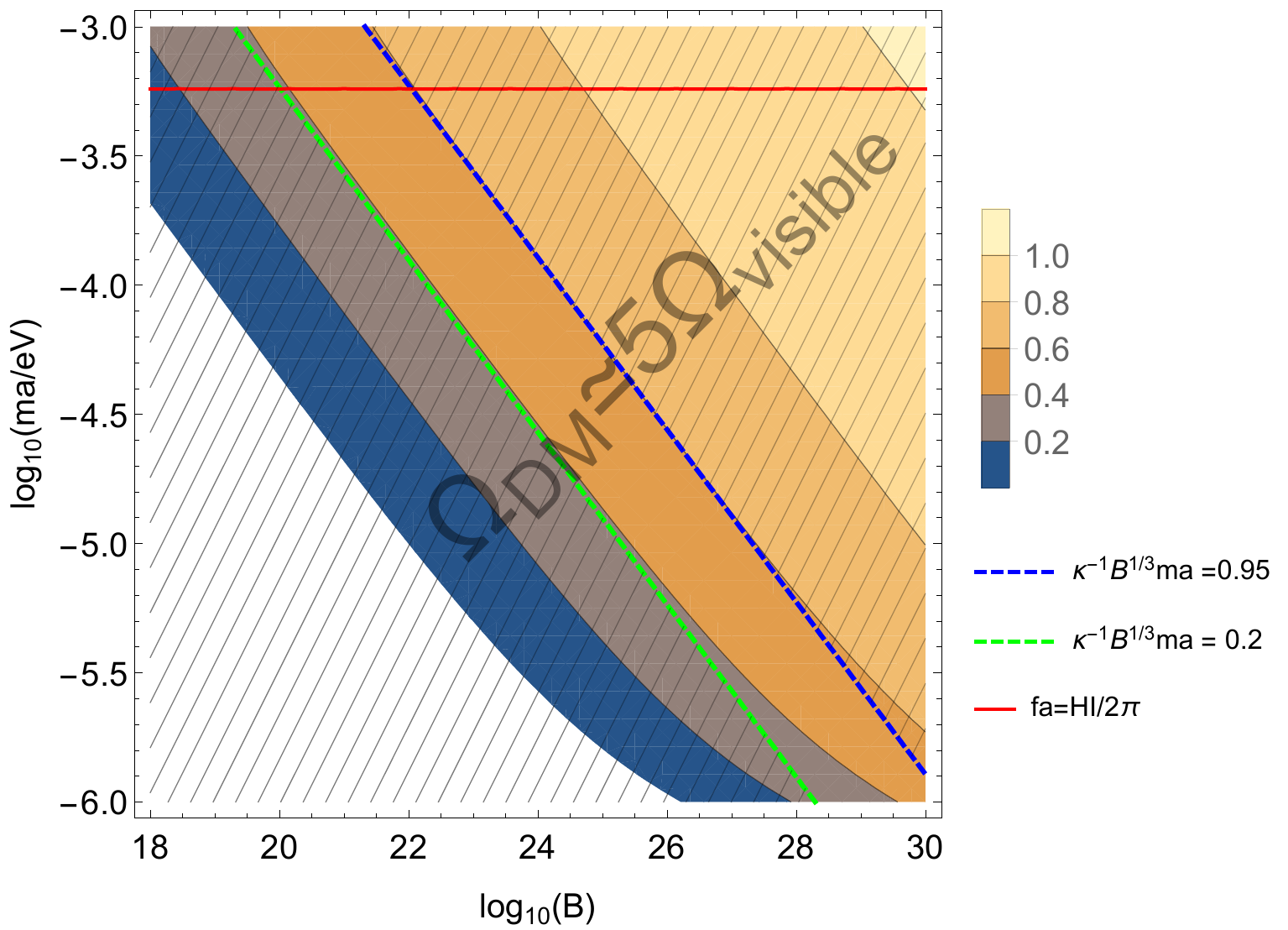}
		\caption{$H_{I}/2\pi=10^{10}~{\rm GeV}, \theta_{a,i}=10^{-1}$}\label{c_HI=2pi10_ntheta=1}
	\end{subfigure}
	\begin{subfigure}[t]{.497\textwidth}
		\includegraphics[width=\textwidth]{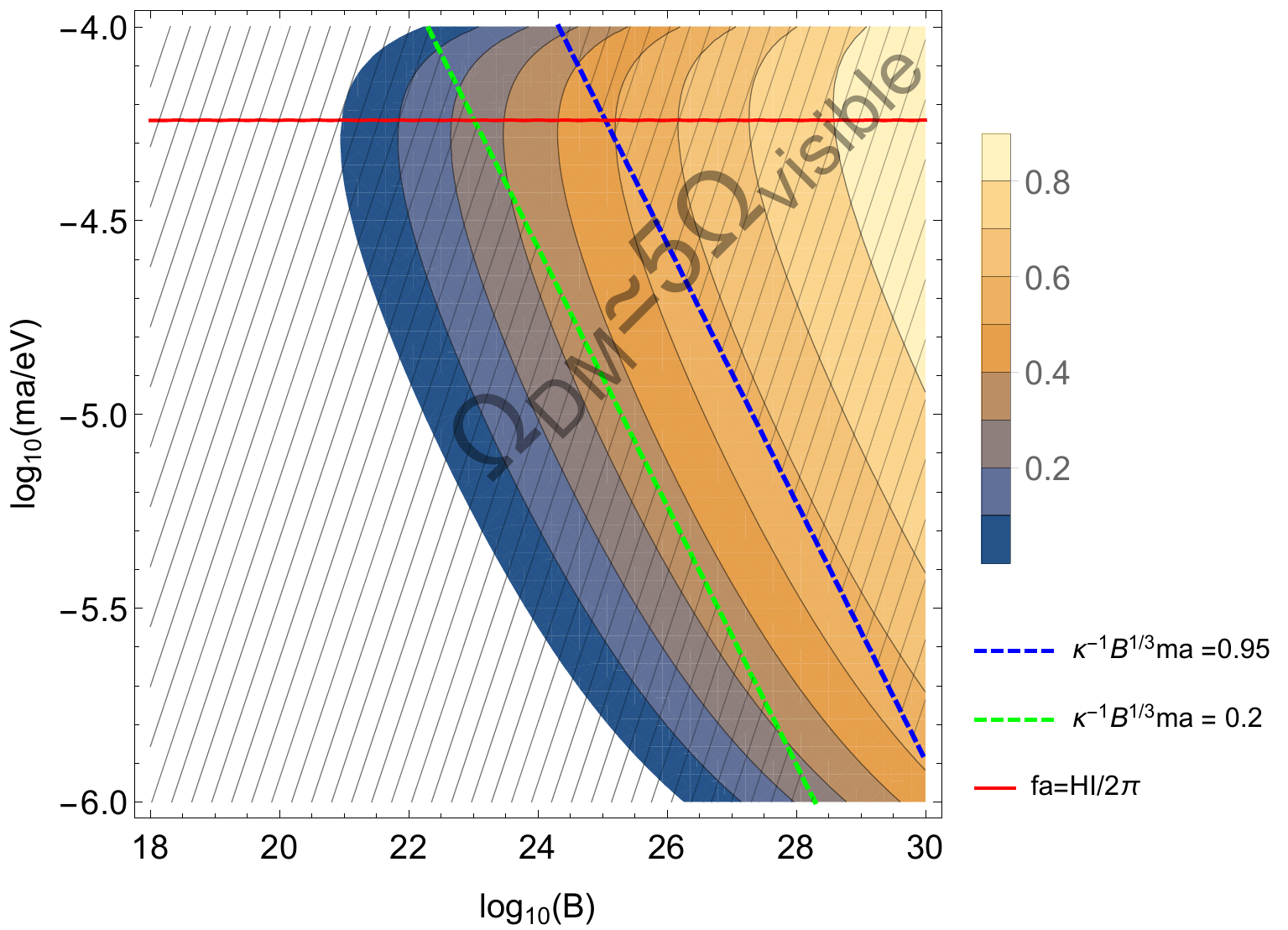}
		\caption{$H_{I}/2\pi=10^{11}~{\rm GeV}, \theta_{a,i}=10^{-1}$}\label{c_HI=2pi11_ntheta=1}
	\end{subfigure}
	\caption{Contour plots of $c$ as a function of $m_{a}$ and $B$ for $H_{I}/2\pi=10^{10}$ GeV and $10^{11}$ GeV respectively with the fixed value $\theta_{a,i}=10^{-1}$ and $\kappa=10^{-4}$. Different colours (shown on the right) correspond to the different values of the parameter $c\sim 1$.}
	\label{fig:plotc2}
\end{figure*}

The key observation here is that there will be always a  region $(B, m_a)$  when the   total dark matter density  assumes its observational value through the parameter $c$ which determines the nugget's  contribution to $\Omega_{\rm DM}$. The corresponding contribution  varies to accommodate the related  axion portion  $ \Omega_a$ as  the total dark matter density $\Omega_{\rm DM}$ on Fig.~\ref{fig:plotc1} is fixed and assumes its observational value. 
 
  \begin{figure*}
	\centering
	\captionsetup{justification=raggedright}
	\begin{subfigure}[t]{.497\textwidth}
		\includegraphics[width=\textwidth]{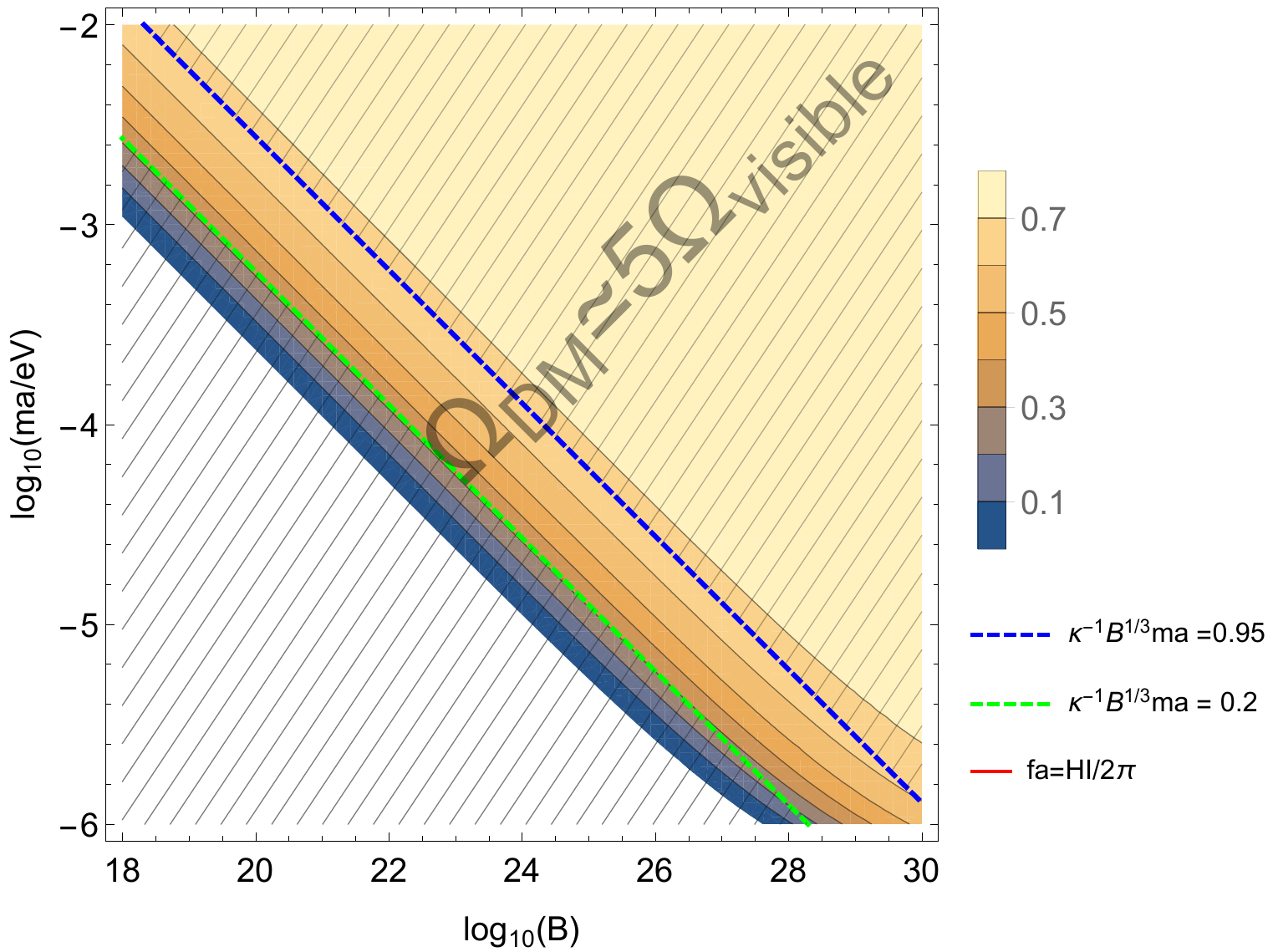}
		\caption{$H_{I}/2\pi=5.7\times10^{8}~{\rm GeV}, \theta_{a,i}=10^{-1}$}\label{model2_M200u400_c_HI=2pi57_ntheta=1}
	\end{subfigure}
	\begin{subfigure}[t]{.497\textwidth}
		\includegraphics[width=\textwidth]{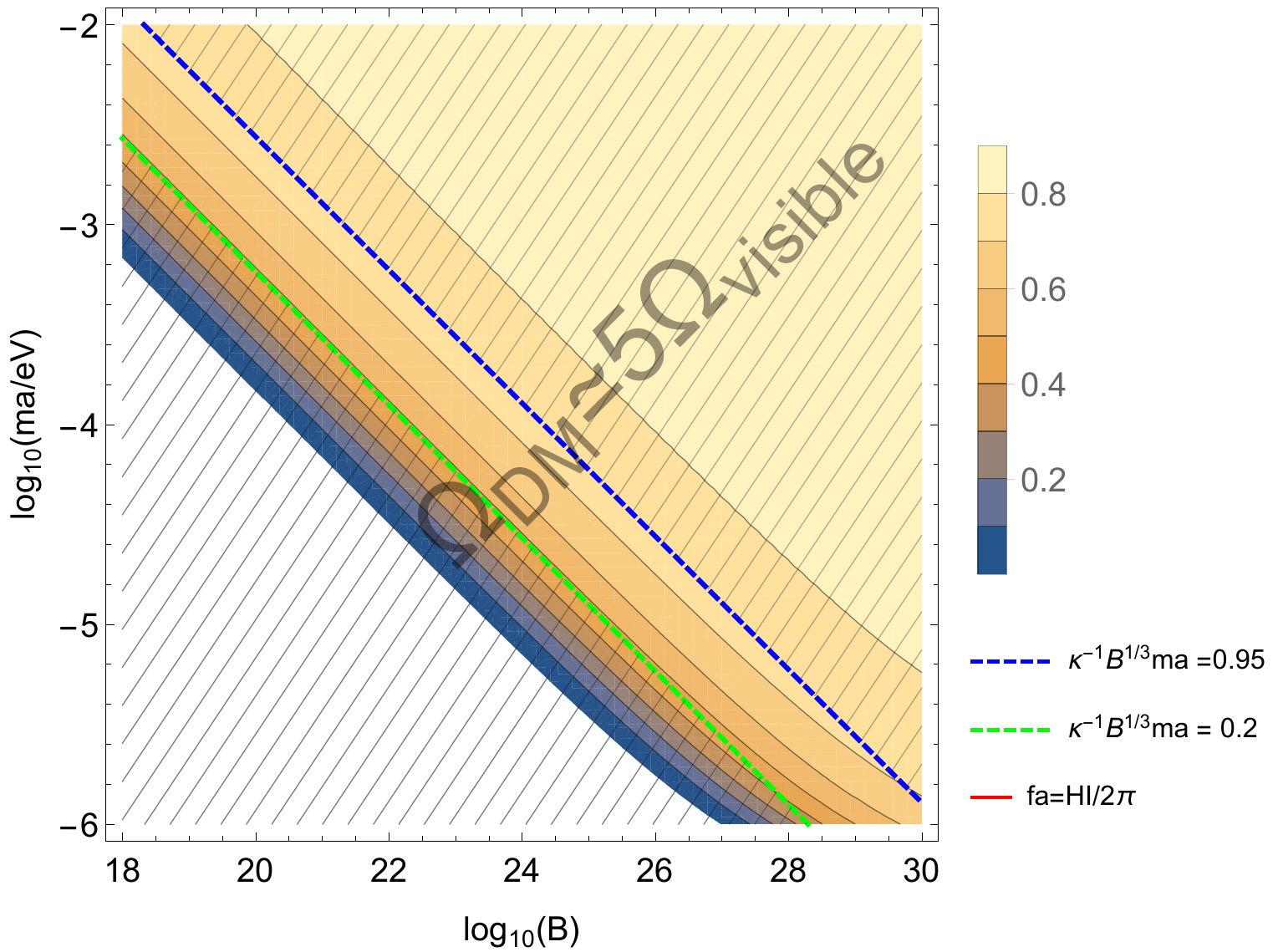}
		\caption{$H_{I}/2\pi=5.7\times10^{8}~{\rm GeV}, \theta_{a,i}=10^{-1}$}\label{model2_M160u500_c_HI=2pi57_ntheta=1}
	\end{subfigure}
	\caption{Model 2: contour plots of $c$ as a function of $m_{a}$ and $B$ for specific values of $H_{I}$,  $\theta_{a,i}$, and   $\kappa=10^{-4}$. Left: $(M_{q},\mu)=(200,400)$~{\rm MeV}. Right: $(M_{q},\mu)=(160,500)$~{\rm MeV}. Different colours (shown on the right) correspond to the different values of the parameter $c\sim 1$.}.
	\label{fig:plotmodel2}
\end{figure*}

 On Fig. \ref{fig:plotc2}  we wish to demonstrate the sensitivity of the allowed region for  region $(B, m_a)$ with respect to $H_I/2\pi$ parameter, 
 where the constraint \eqref{eq:HI_constraint} is applied. 

 We want to demonstrate the corresponding sensitivity to $H_I/2\pi$ parameter   by showing that 
 there is no any dependence on $H_I$ for sufficiently small $H_I/2\pi\simeq  10^{10}~{\rm GeV}$  as shown on Fig. \ref{c_HI=2pi10_ntheta=1}. 
 In all respects the plot  is very much the same as the one shown in Fig.~\ref{c_HI=2pi57_ntheta=3}.  In both  cases the dark matter density is  dominated by the nuggets, and the the allowed region $(B, m_a)$ is  not sensitive to the $H_I/2\pi$ as long as the Hubble parameter  is sufficiently small. 
 However,  when $H_I/2\pi$ becomes close to $f_a \sim 10^{11}$~ GeV, the window for $c$  is shifted as shown on Fig. \ref{c_HI=2pi11_ntheta=1} to accommodate the conventional contribution of the propagating axions.  The main point is that  there will be always a  region $(B, m_a)$  when the   total dark matter density  assumes its observational value, though the magnitude of  $c\in (0, 1)$ assumes somewhat different values, depending on external  
  parameters of the system.

Our next task is to analyze  the sensitivity of our results to the QCD parameters related to CS properties of the nuggets. To accomplish this goal 
we plot parameter $c$ on Fig.~\ref{fig:plotmodel2}  as a function of $m_{a}$ and $B$ using $\epsilon_{\rm tot}^{(2)}$ for Model-2 determined by eqs.(\ref{eq:3.Model 2 epsilon}) and (\ref{eq:3.Model 2 M mu}).
The corresponding plot for  $(M_{q},\mu)=(200, 400)~{\rm MeV}$ is presented on   Fig.~\ref{model2_M200u400_c_HI=2pi57_ntheta=1} while  for  $(160,500)~{\rm MeV}$ is shown on   Fig.~\ref{model2_M160u500_c_HI=2pi57_ntheta=1} respectively. 

The main conclusion is that the Model-2 (which is based  on the fundamentally different building principles than Model-1) with varies  parameters produces  nevertheless quantitatively similar results as Model-1 analyzed previously  and shown on Fig. \ref{c_HI=2pi57_ntheta=3}. 
A more detail comparison presented in  Appendix \ref{appendix:A.model comparison} supports this claim. This conclusion  essentially implies  that 
our phenomenological results are not very sensitive to the specifics of the QCD parametrization of the system  describing the dense CS phase of matter in strongly coupled regime. Therefore, we treat our results as the solid consequences of the AQN model.

As an additional note, the parameter $c_\Omega$ as defined by eqs.(\ref{eq:c_Omega}) and (\ref{eq:c_relation}) 
and which describes the mass-difference  between the nuggets and antinuggets (in contrast with parameter $c$ which describes the baryon charge-difference between the nuggets and antinuggets) is numerically very close to parameter  $c$ studied above. Specifically, one can show that for  Model-1 \eqref{eq:3.Model 1 epsilon_final} the parameter  $c_\Omega\simeq1.17c$ within  $15\%$ accuracy for the  region  $0.4\lesssim c\lesssim0.6$ which dominates the parametrically allowed region    as discussed  in the preceding paragraphs. Therefore, we do not show the plots for  $c_\Omega$ 
as a function of external parameters because they are  very similar to the plots for $c$ presented and discussed above.

\subsection{No fine-tuning   in the AQN scenario}\label{subsec:tuning}
\begin{figure}
	\centering
	\captionsetup{justification=raggedright}
	\includegraphics[width=.497\textwidth]{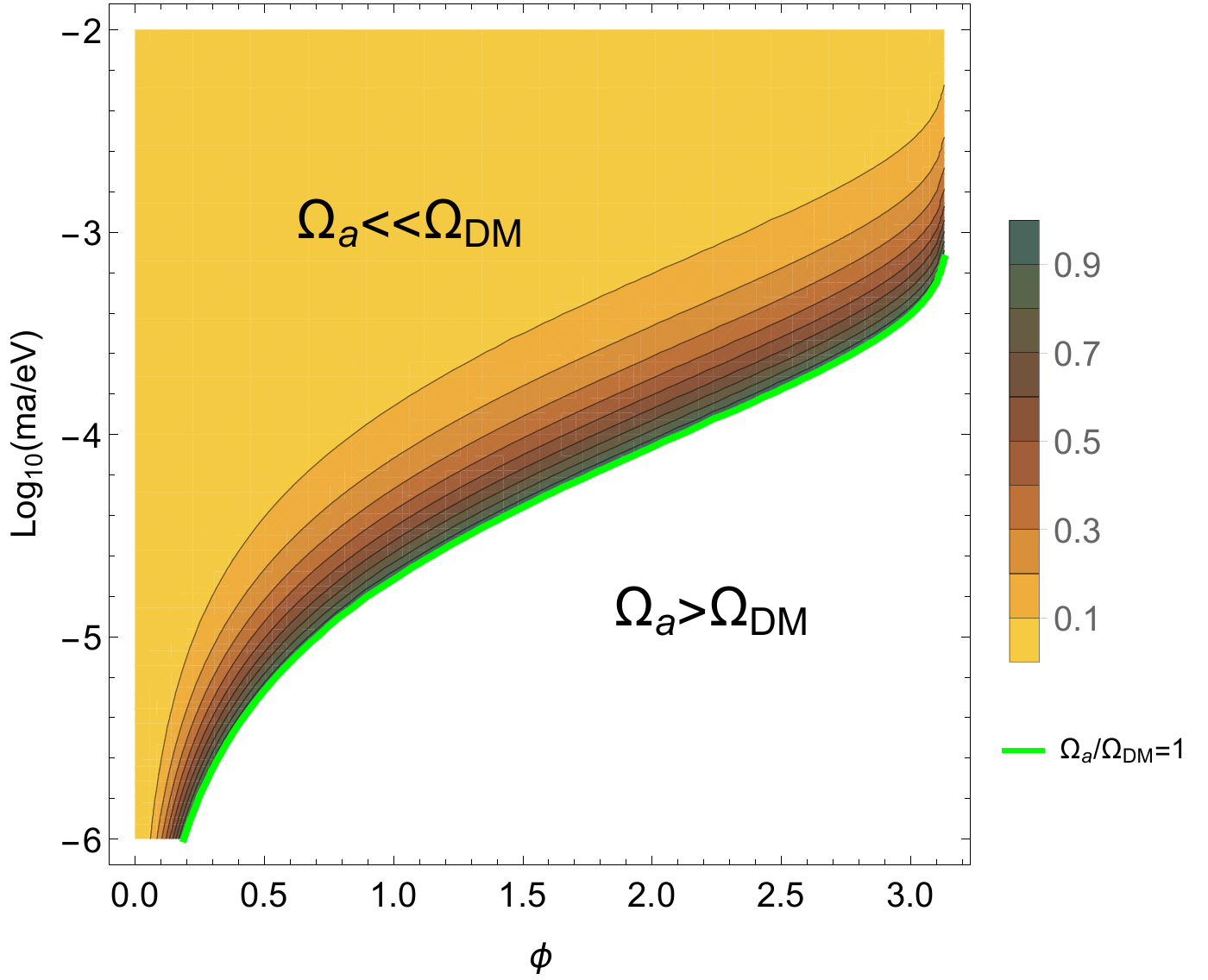}
	\caption{Contour plots of $\Omega_{a}/\Omega_{DM}$ as a function of $m_{a}$ and $\phi=\sqrt{\theta_{a,i}^{2}+\left(H_{I}/(2\pi f_{a})\right)^{2}}$.
	Different colours (shown on the left) correspond to the different values of $\Omega_{a}/\Omega_{DM}$.}
	\label{fig:nonsensitivity_Omegaa}
\end{figure}

As we mentioned in section \ref{sec:2.density}, the dark matter propagating  axion itself may not   saturate the total dark matter.
In contrast, the nugget's  formation  always generates a large  contribution $\Omega_{\rm DM}\sim \Omega_{\rm visible}$ and always accompanies  the conventional axion production. 
This property of the AQN model is demonstrated  on 
 Fig.~\ref{fig:nonsensitivity_Omegaa}, where we plot $\Omega_{a}/\Omega_{DM}$ as a function of $m_{a}$ and $\phi=\sqrt{\theta_{a,i}^{2}+\left(H_{I}/(2\pi f_{a})\right)^{2}}$. The function $\phi(\theta_{a,i},H_{I})$ enters formula (\ref{eq:Omega_a_full}) for $\Omega_a$  and counts together with the initial homogeneous displacement contribution and the backreaction contribution to the free dark matter axions. The  Fig.~\ref{fig:nonsensitivity_Omegaa} 
 explicitly shows that  that $m_{a}$ and $\phi$ have to be highly fine tuned   to make $\Omega_{a}$ to  saturate $\Omega_{DM}$ exactly, shown as a bright green solid line. 
 In other words, for a specific  magnitude  of $m_a$ there is a single  value of $\phi$ when the dark matter density assumes its observable value. 
 Once these two parameters,  $m_a$  and $\phi$ slightly deviate   from the appropriate values, the  $\Omega_{a}$ strongly deviates from    $\Omega_{DM}$. 
 
 This conventional fine- tuning scenario should be contrasted with the results of the AQN model when 
 $\Omega_{a}$ may contribute very little to  $\Omega_{DM}$. Nevertheless, the $\Omega_{DM}$ assumes its observation value as a result of an additional nugget's contribution  which always accompanies  the  axion production and always generates a contribution of order one, as we already emphasized. In other words, 
  the AQNs   play the role of the ``remaining'' DM density which, in fact, could be the dominating portion of the $\Omega_{DM}$. As we have seen in previous subsection \ref{subsec:plots}, there will always be contribution  to the DM model constituted of free axions and the AQNs for the allowed parametric  space.  In other words, for a specific  magnitude  of $m_a$ there is a large window of $\phi$ corresponding to the different values of parameter $c\in (0,1)$   when the dark matter density assumes its observable value. Different colours on the Fig.~\ref{fig:nonsensitivity_Omegaa}
  correspond to the different axion contribution   represented by parameter $\Omega_a/\Omega_{\rm DM}$.
  Therefore,     the fine-tuning problem does not even occur in the AQN scenario as entire parametrical space on the left from the green solid curve $\Omega_a=\Omega_{\rm DM}$ corresponds to the observable dark matter density.  The white region in Fig.~\ref{fig:nonsensitivity_Omegaa} is excluded  because   the requirement $\Omega_{a}\leq\Omega_{\rm DM}$ is violated. 

\begin{figure}
	\centering
	\captionsetup{justification=raggedright}
	\includegraphics[width=.497\textwidth]{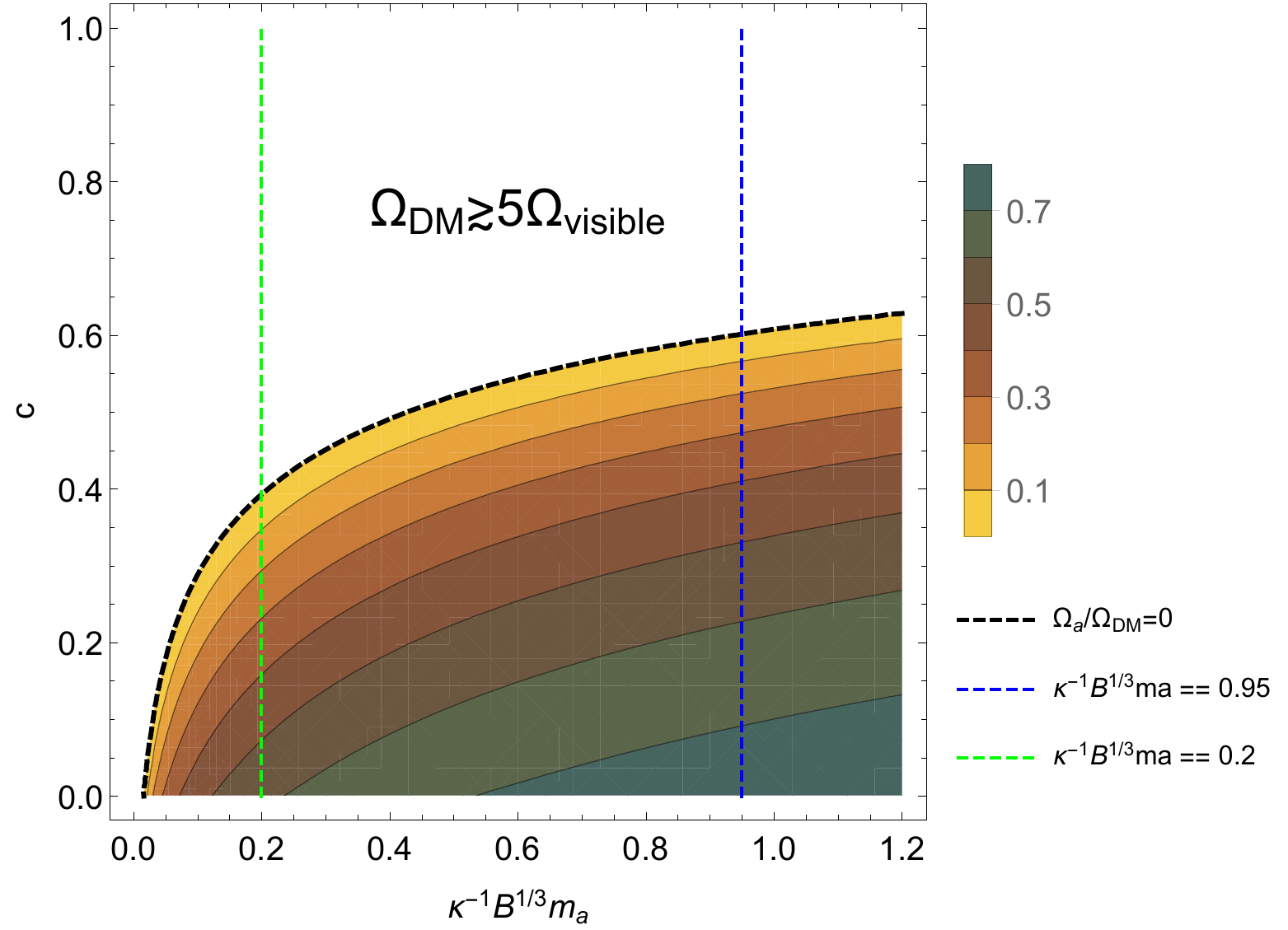}
	\caption{Contour plots of $\Omega_{a}/\Omega_{DM}$ as a function of $c$ and $\kappa^{-1}B^{1/3}m_{a}$. Different colours (shown on the right) correspond to the different values of $\Omega_{a}/\Omega_{DM}$. }
	\label{fig:nonsensitivity_fitting}
\end{figure}

As the final technical remark we also notice from Fig.~\ref{fig:nonsensitivity_Omegaa} that for most part of the parameter space, we have $\Omega_{a}\lesssim0.2\Omega_{\rm DM}$.
To make the above statement more precise, we plot on Fig.~\ref{fig:nonsensitivity_fitting} the ratio $\Omega_{a}/\Omega_{\rm DM}$ as a function of $c$ and parameter $\kappa^{-1}B^{1/3}m_{a}/m_\pi$ determined by the QCD physics as given by eq.\eqref{eq:3.Model 1 epsilon_final}.  The white region in Fig.~\ref{fig:nonsensitivity_fitting} stands for the excluded region of parameters   ($c$, $\kappa^{-1}B^{1/3}m_{a}$). This plot shows that 
   the parameter $c$ cannot be very close to $\sim$1 for the allowed QCD window  $0.2\lesssim\kappa^{-1}B^{1/3}m_a/m_\pi\lesssim0.95$. This property, in fact,  can be understood analytically  from eq.(\ref{eq:c_previous}) or its generalized version eq. (\ref{eq:results}) where $c\rightarrow 1$ implies $\Omega_{\rm DM}\gg\Omega_{\rm visible}$ which violates the observable relation $\Omega_{\rm DM}\simeq5\Omega_{\rm visible}$. The main conclusion to be drawn   from this plot is that  the parametrical region where $\Omega_{DM}$ assumes its observable value is
very large and perfectly consistent with the QCD constraints related to parameters $c$ and  $ \kappa^{-1}B^{1/3}m_{a}/m_\pi$.

This conclusion is another  manifestation of the basic consequence of the AQN framework when the relation (\ref{Omega}) is not sensitive to any details of the system such as $m_a$ or $\theta_{a,i} $, but rather represents a direct  outcome of this   proposal.
This fundamental result is essentially incorporated into the initial building principle of the entire framework, and cannot disappeared as a result of  some additional technical details and modifications.

\section{Conclusion and future development}
\label{sec:5.conclusion}

This work is the first attempt to extract  the key element of the proposal, the coefficient  $``c"$ describing the disparity    between nuggets and antinuggets, as defined by eq.(\ref{eq:c_previous}) or its generalized version eq.(\ref{eq:results}), 
  from the observational constraints in the quantitative way.
Precisely this asymmetry eventually determines the dark matter density (within this framework) we observe today as a result of the charge separation mechanism replacing the conventional ``baryogenesis" scenario as overviewed  in section \ref{intro}. This  work, to a large extend,      is  motivated
by the recent progress in the axion search  experiments which, hopefully,  in few years should   cover  almost entire interesting region of the allowed parametrical space for $m_a$ as recently reviewed in
\cite{Rosenberg:2015kxa,Marsh:2015xka,Graham:2015ouw,Ringwald2016}. 

A distinct feature of the AQN model is that the fundamental relation (\ref{Omega}) is always satisfied in this framework  irrespectively to the parameters of the system, such as the axion mass $m_a$  or misalignment angle $\theta_{a,i}$.
Therefore, the discovery  of the axion with mass $m_a$ which gives  only a fraction  of the observed dark matter density $\Omega_a< \Omega_{\rm DM}$ is perfectly consistent with our scenario as the remaining portion of the dark matter is generated  by the nugget's contribution (\ref{eq:ratio}) 
which always accompanies the axion production within  this proposal. 
 
 The corresponding numerical results are presented in section \ref{subsec:plots} which explicitly demonstrate that 
 this model is perfectly consistent with varies constraints  listed in section \ref{subsec:constraints}.  One should emphasize that  the corresponding constraints have been derived from a variety of   astrophysical, cosmological, satellite and ground based observations.
Furthermore, there is a number of frequency bands where some excess of emission was observed, but not explained by conventional astrophysical sources. This model may explain some portion, or even entire excess of the observed radiation in these frequency bands, see short review \cite{Lawson:2013bya}.   This model is   also consistent with known constraints from the axion search experiments. 
 Finally,  in section \ref{subsec:tuning} we argued that the corresponding results are not very sensitive to varies QCD parameters describing the dense CS phase of the AQNs. 
 
 Therefore, there is no much sense  to repeat here, in conclusion, the  main  results of the sections \ref{subsec:constraints},  \ref{subsec:plots},  \ref{subsec:tuning}. Rather, we would like to discuss the implications of these results on  the future related studies.
  
 In particular, it has been recently argued that the dark matter axions can generate peculiar effects in the so-called Josephson junctions
 \cite{Beck:2013jha,Beck:2014aqa,Beck:2017sle}. If a small measured peak of unknown origin is identified with the dark matter axions as suggested
 in \cite{Beck:2013jha,Beck:2014aqa,Beck:2017sle}, one can infer that the axion mass
 should be 
 \be
 \label{Josephson}
 m_a=(104.0\pm 1.3)\cdot 10^{-6} {\rm eV},~~~ \rho_a\simeq 0.1 ~ \frac{{\rm GeV}}{\rm cm^3},
 \ee
  where we also quote the estimated  axion dark matter density $\rho_a$ which  was  required to  interpret  the measured signal  in the Josephson junctions  as a result of the DM axion \cite{Beck:2017sle}.  This  value of $\rho_a$   represents approximately a quarter of the expected dark matter density in the halo. One should emphasize that 
  the estimate  (\ref{Josephson}) for the axion mass was extracted from analysis of four different experiments observed by four different groups pointing to the same mass  value (\ref{Josephson}).
  Therefore, it is very unlikely that (\ref{Josephson}) is  a statistical fluke. 
  
  If one  literally accepts  the measured signal  in the Josephson junctions  as a result of the DM axion  (\ref{Josephson}) 
  one  could immediately infer, for example,  from Fig.\ref{fig:nonsensitivity_Omegaa} that for $m_a\simeq 10^{-4} {\rm eV}$ and $\Omega_a/\Omega_{\rm DM} \simeq 0.3$ as eq. (\ref{Josephson})  would suggest,  the misalignment angle $\theta_{a,i}$ should be order of  one provided  that $\left(H_{I}/2\pi f_{a}\right)\ll 1$. Furthermore, from Fig. \ref{c_HI=2pi57_ntheta=0} for the axion mass $m_a\simeq 10^{-4} {\rm eV}$  and $\theta_{a,i}\simeq 1$ one could estimate the average baryon charge of the AQNs which is estimated to be  order of $B\sim 10^{25}$. This  estimate  is very encouraging as it    is perfectly consistent with our previous phenomenological analysis reviewed in section \ref{sec:AQN}. 
  
 One should emphasize that such kind of estimates and self-consistency checks   are very important for  subsequent  development and future studies. For example,  in ref.~\cite{Zhitnitsky:2017rop} it has been argued that  the Sun could serve as an ideal lab to study the AQN model. The main point is that the antinuggets  deposit   a huge amount of  energy in the corona as a result of annihilation events  with the solar material.  It may explain  the so-called ``the Solar Corona Mystery" when the temperature of the corona is about  $10^{6}$~K, i.e., being a few 100 times hotter than the solar surface temperature. It may also explain the extreme UV and soft x-ray emissions from corona, which is very hard to  explain using conventional astrophysical processes.  
It has been also  suggested   in ref.~\cite{Zhitnitsky:2017rop} that the observed  brightening-like  events called  ``nanoflares" in the Sun can be identified with  the annihilation  events of the antinuggets, in which case the observed energy distribution of the nanoflares must coincide with the baryon charge   distribution  studied in  the present work.  In other words, the energy distribution of the nanoflares and the baryon charge distribution of the nuggets is one and the same function within the AQN model ~\cite{Zhitnitsky:2017rop}. 

   This statement can be formally expressed as follows  
  \be
  \label{distribution}
   {dN} \sim B^{-\alpha}dB\sim W^{-\alpha} dW, 
  \ee
  where $dN$ is the number of the nanoflare  events per unit time with energy between $W$ and $W+dW$ which occur as a result of complete annihilation 
  of the antinuggets carrying the  baryon charges between  $B$ and $B+dB$.
 These two distributions are tightly linked  
    as these two entities are related to the same AQN objects when the  annihilation of the antinugget's  baryon charge   generates the energy
    which is    interpreted as  the observed nanoflare event within AQN model. The modelling and observations of the nanoflares in corona suggest $\alpha\sim 2$, see ~\cite{Zhitnitsky:2017rop} with the details and references.  One should emphasize that this interpretation is consistent
    with constraints on $B$, see (\ref{B-range}),  refs.  \cite{Lawson:2013bya, Jacobs:2014yca} for review and   footnote \ref{B-constraint} with some additional  remarks. Therefore, further studies of the nanoflares in solar corona may shed some light on  the nature of the dark matter.

In fact this interpretation may   receive further support  by future  analysis   of some specific  correlations studied in \cite{Zioutas}. 
If subsequent development  along the lines advocated in  \cite{Zioutas} suggesting that the frequency of the flares observed in the sun is  directly related to the dark matter particles (called ``invisible matter"  in ref. \cite{Zioutas}), it would be a major breakthrough not only in our understanding of the solar corona, but also in our understanding of the nature of the dark matter. 

The first theoretical  estimates  providing a specific mechanism on how the correlations observed  in \cite{Zioutas}
could be, in principle, explained  have  been recently suggested  in   \cite{Zhitnitsky:2018mav}.
The basic idea  of \cite{Zhitnitsky:2018mav} is that the nuggets entering the solar corona will inevitably generate the shock waves
as the typical velocities of the nuggets are much higher that the speed of sound in solar atmosphere. The shock waves due to the AQNs may serve as the triggers igniting  the large flares which are shown to be correlated with ``invisible matter" of ref. \cite{Zioutas}. 
This mechanism would also   relate naively unrelated entities such as  the axion, its mass,     the baryon charge distribution $B$ of the nuggets (\ref{distribution}),   the flare's  intensity and their frequency of appearance   in the Sun.

Our next comment about possible future studies is related to recent activities of the axion search experiments. 
 We want to specifically mention some  present and future axion search experiments such as ADMX and ADMX-HF \cite{ADMX},  IAXO \cite{Armengaud:2014gea},  CAST \cite{Anastassopoulos:2017ftl}, ORPHEUS  \cite{Rybka:2014cya}, 
MADMAX \cite{TheMADMAXWorkingGroup:2016hpc}, ORGAN  \cite{McAllister:2017lkb}. These, and possibly many other experiments  should eventually discover the QCD axion
irrespectively to the assumption on validity of the axionic Josephson effect leading to (\ref{Josephson}). The only original comment 
we would like to make  is that  the discovery  of the axion with mass $m_a$ which may generate   only a small fraction  of the observed dark matter density $\Omega_a\ll  \Omega_{\rm DM}$ nevertheless would be a major discovery  because the remaining portion of the dark matter could be  generated  by the nugget's contribution (\ref{eq:ratio}) 
which always accompanies the conventional axion production within  the AQN scenario, and always satisfies the generic relation (\ref{Omega}). 
 This would conclude a long and fascinating journey of searches  for this unique and amazing particle conjectured almost 40 years ago.
 
We want to make one more comment   on the axion search experiments when the observable is sensitive to the axion amplitude $\theta$ itself, in contrast with conventional proposals which are normally sensitive to the derivatives $\sim\partial_{\mu}{\theta}$.     The basic idea of the proposal \cite{Cao:2017ocv}   is that the  $\theta$  becomes a physically  observable parameter even in the abelian Maxwell QED if the system is defined in a topologically nontrivial sector. This can be easily   achieved by  placing the system into the background of an external magnetic field. 
The phenomenon  in all respects is very similar to well known Witten's effect when   the $\theta$ parameter becomes a physical observable  in the presence of the magnetic monopole.  This novel phenomena was coined as the  Topological Casimir Effect (TCE), and there are some specific
 ideas how 
to design and fabricate  the corresponding apparatus   (the so-called aKWISP project), see talk by Cantatore \cite{Cantatore} with relevant information.

One more possible  direction for future studies from our ``wish list'' is a development of the QCD-based technique related to nuggets evolution, cooling rates, evaporation rates, viscosity, transmission/reflection coefficients, etc., in an unfriendly environment with non-vanishing $T, \mu, \theta$.  The problem here is that the axion mass scale $m_a$ and the Hubble scale $H(T)$ at the QCD epoch are  drastically different from the QCD scale itself $\sim \Lambda_{\rm QCD}$ describing  the CS phase and the nugget's structure. In similar  circumstances some  researches normally change  the scales (in which case  the relevant parameters obviously assume some unphysical values)  to attack the problem numerically. After the numerical computations are done, one can return to the physically relevant  values for the parameters by using   plausible arguments with some specific  assumptions
(which may or may not be correct) on the scaling features of  the parameters   when they  assume their physical values. 

Our   goal in  the nearest future is to develop some numerical methods and approaches  to study  the real time evolution of nuggets with real physical parameters which assume drastically different scales.
If the project turns out to be successful  it would  be a major technical step forward relating the  analysis of refs \cite{Liang:2016tqc,Ge:2017ttc}  which was devoted to  the formation period of the nuggets at high temperature and present work which is dealing with the  present epoch of the cold Universe, billion years  after the nuggets had been formed.

\section*{Acknowledgments}

This work was supported in part by the National Science and Engineering
Research Council of Canada.

\appendix

\section{Comparison of the  models}
\label{appendix:A.model comparison}
\begin{figure}
	\centering
	\includegraphics[width=\linewidth]{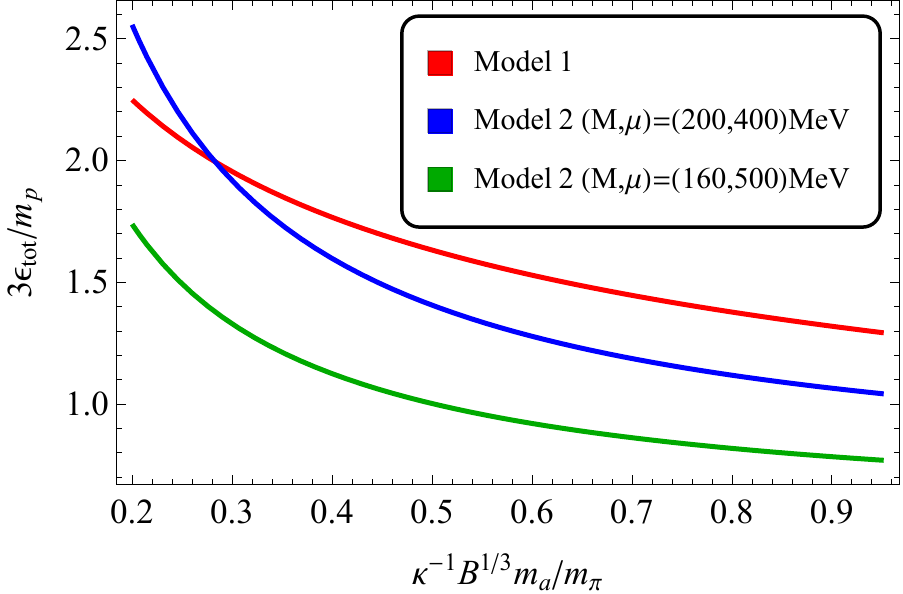}
	\caption{Plot of  $3\epsilon_{\rm tot}/m_p$ vs. $\kappa^{-1}B^{1/3}m_a/m_\pi$.}
	\label{fig:epsilonmptot}
\end{figure}

Our goal here is to overview different QCD- inspired models which we used in our estimates for the energy per baryon charge in  CS phase in   Sec. \ref{sec:3.epsilon model}.
 First, we note that both models, the Model-1 [Eq. \eqref{eq:3.Model 1 epsilon_final}] and Model-2 [Eq. \eqref{eq:3.Model 2 epsilon}] are sensitive to a single dimensionless    parameter $\kappa^{-1}B^{1/3}m_a/m_\pi$. In fact, the two models give similar curves of total energy per baryon charge (which includes the energy related to the axion domain wall $\epsilon_{\rm DW}$ as well as Fermi energy of the quarks $\epsilon_{\rm QCD}$). It is quite a remarkable result as the models are based on fundamentally different building principles. 
 
 The results are    shown on Fig. \ref{fig:epsilonmptot}, where we have rescaled it by  hadronic energy density $\frac{1}{3}m_p$ per baryon charge. The  short conclusion is as follows. Although the two models are build from  fundamentally different principles, they turn out to give similar results. Such close agreement supports our estimations on the energy density of a stabilized  quark nugget we used in the main text in Section \ref{sec:3.epsilon model}.
 \begin{figure}
	\centering
	\includegraphics[width=\linewidth]{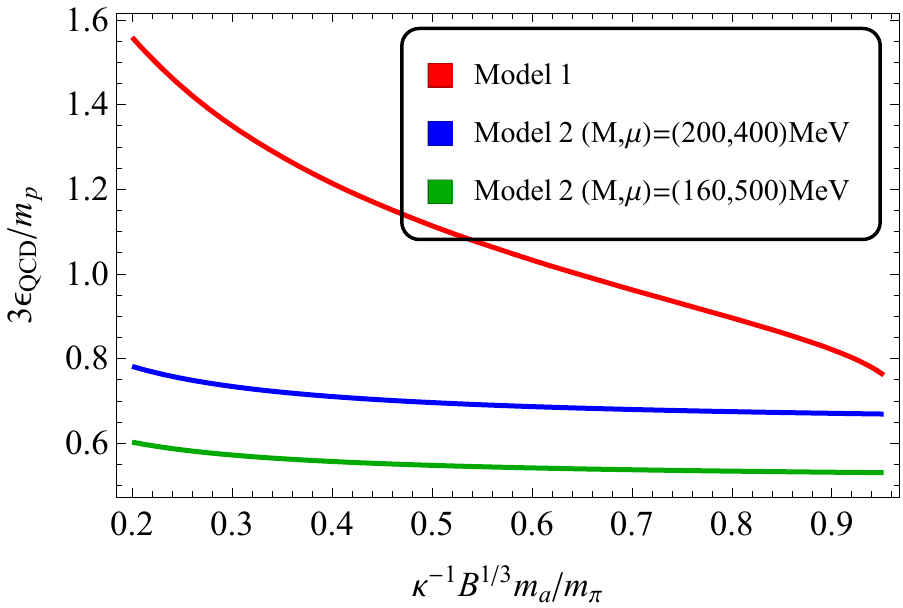}
	\caption{Plot of $3\epsilon_{\rm QCD}/m_p$ vs. $\kappa^{-1}B^{1/3}m_a/m_\pi$.}
	\label{fig:epsilonmpqcd}
\end{figure}

\begin{figure}
	\centering
	\includegraphics[width=\linewidth]{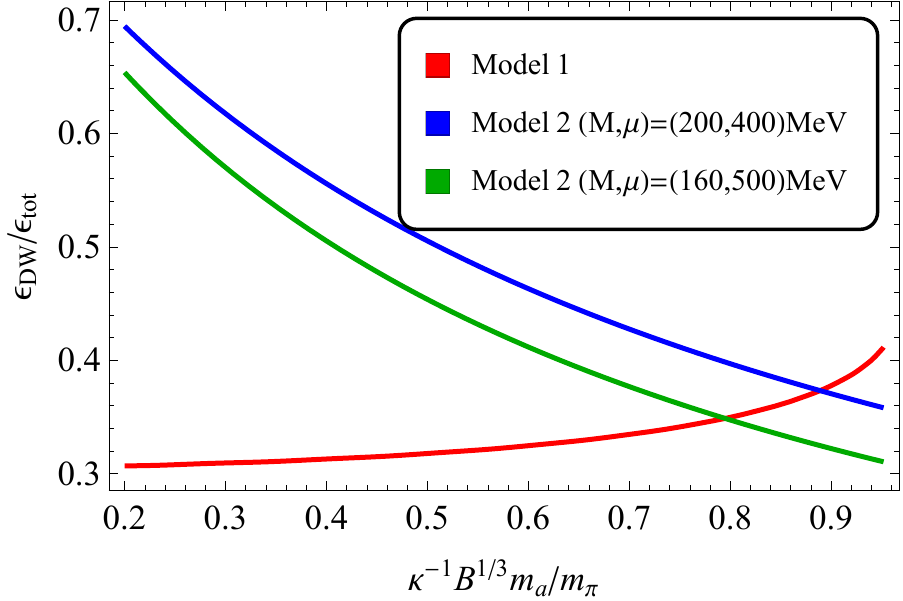}
	\caption{Plot of $\epsilon_{\rm DW}/\epsilon_{\rm tot}$ vs. $\kappa^{-1}B^{1/3}m_a/m_\pi$.}
	\label{fig:epsilonmpwall}
\end{figure}

  In addition to this main conclusion we want to make some additional comments below. In the plot   shown on Fig.\ref{fig:epsilonmptot} we do not extend the analysis beyond the region when parameter  $\kappa^{-1}B^{1/3}m_a/m_\pi$ becomes sufficiently large. The main reason to limit ourselves by this window is that  the larger values of the parameter $\kappa^{-1}B^{1/3}m_a/m_\pi >0.95 $  correspond to smaller chemical potential\footnote{\label{mu}More specifically, $\mu$ can be evaluated from Eq. \eqref{eq:3.Model 1 x sigma}, where $x_{\rm eq}$ is numerically solved as a function of $\kappa^{-1}B^{1/3}m_a/m_\pi$ from condition \eqref{eq:3.Model 1 equilibrium}. 
Large value of  $\kappa^{-1}B^{1/3}m_a/m_\pi>0.95 $  corresponds  to the small chemical potential $\mu\lesssim330$ MeV. At this relatively small value of $\mu$,  our treatment of  the dense matter  as the  colour superconductor is not justified. The reason is that $\mu\lesssim330$ MeV
  corresponds to the conventional stable nuclear matter which should be treated differently as the   relevant degrees of freedom are   hadrons, rather than quarks.}. For the same reason, only plots in region $\kappa^{-1}B^{1/3}m_a/m_\pi\leq0.95$ has been  considered in this work. 
  
 This constraint does not imply that very  large nuggets cannot exist. In fact the quark nuggets can perfectly coexist with conventional nuclear matter
 making an absolutely stable object, when the hadronic nuclear matter is surrounding the dense  the quark nugget.
Such objects could be captured by  stars or planets and stay in the cores of the astronomical object indefinitely\footnote{In fact, there are many arguments suggesting  that the cores of the neutron stars could be   in CS phase.}.
 They can also accrete the visible hadronic matter during a long Hubble evolution since the formation time at $T\simeq 41$ MeV as shown on Fig.\ref{phase_diagram}. However, an antinugget made of antimatter will annihilate its baryon charges  when  it is  in   contact with visible matter.
  As the main phenomenological manifestation of the AQN model is precisely such annihilation events of the antinuggets with visible matter 
  in the present work we present our plots for the antinuggets and their baryon charges $B$.

 The absolute stability of the AQNs is determined by the condition $\frac{3\epsilon_{\rm QCD}}{m_p}\leq1$, while the metastability (with very large life time exceeding the life time of the Universe) this condition may not be strictly satisfied. The corresponding plots are shown 
 on Fig. \ref{fig:epsilonmpqcd}. The main lesson from these studies is that the Model-2  is absolutely stable in the entire  parametrical region    $0.2\lesssim\kappa^{-1}B^{1/3}m_a/m_\pi\lesssim0.95$, while Model-1 is absolutely stable for $0.6\lesssim\kappa^{-1}B^{1/3}m_a/m_\pi\lesssim0.95$,
and  it becomes metastable for $0.2\lesssim\kappa^{-1}B^{1/3}m_a/m_\pi\lesssim0.6$. This metastability (in contrast with absolute stability) should not be a point of concern as discussed in the original  paper \cite{Zhitnitsky:2002qa} because it  corresponds to a very long life time of the nuggets.
This metastability region  can be ignored for any practical purposes in our discussions in the present work. 

 For illustrative purposes we also present on Fig. \ref{fig:epsilonmpwall} the plot for the domain wall contribution $\epsilon_{\rm DW}/\epsilon_{\rm tot}$ as a function of the same dimensionless parameter  $ \kappa^{-1}B^{1/3}m_a/m_\pi $. One can explicitly see from this plot that the energy related to the axion field   represents a  considerable portion of the nugget's total energy
 representing approximately 1/3 of the total energy for the Model-1 (red curve). However, this contribution is very distinct from the conventional propagating axions (represented by $\Omega_a$ in this work) which are produced as a result of misalignment mechanism. The energy of the axion field represented by  $\epsilon_{\rm DW}/\epsilon_{\rm tot}$  cannot be easily released to space as the axions describing the axion domain wall are not on-shell axions. The corresponding energy in form of the propagating axions can be only released to the space when the nuggets get completely annihilated and the axions released, for example in the solar corona as discussed in  ~\cite{Zhitnitsky:2017rop,Zhitnitsky:2018mav}. It would be a major discovery if these axions can be observed by  CAST \cite{Anastassopoulos:2017ftl} or IAXO \cite{Armengaud:2014gea} type instrument.

%
%
%


\end{document}